\documentclass[twocolumn]{aa}
\usepackage[varg]{txfonts}

\usepackage{graphicx}
\usepackage{natbib}
\usepackage{color}
\usepackage{subfig}
\usepackage[switch]{lineno}
\usepackage{xspace}
\usepackage{microtype}
\usepackage{amssymb}
\usepackage{mathtools}
\usepackage{multirow}
\def\link_col{blue}
\usepackage{xspace}
\usepackage{url}
\usepackage{wasysym}
\usepackage{caption}
\usepackage[version=4]{mhchem}

\def\gray{$\gamma$-ray\xspace}
\def\grays{$\gamma$ rays\xspace}
\def\fermi{{\it Fermi}-LAT\xspace}

\def\deg{\hbox{$^\circ$}}

\begin{document}

\title{A hard spectrum diffuse $\gamma$-ray component associated with \ion{H}{ii} gas in the Galactic plane }
\titlerunning{$\gamma$-ray component associated with \ion{H}{ii} gas}
    \author{Bing Liu\inst{1,2,3}
    \and Rui-zhi Yang\inst{1,2,3}
    }
    \institute{Department of Astronomy, School of Physical Sciences, University of Science and Technology of China, Hefei, Anhui 230026, China. e-mail: yangrz@ustc.edu.cn
    \and  CAS Key Laboratory for Research in Galaxies and Cosmology, University of Science and Technology of China, Hefei, Anhui 230026, China 
    \and School of Astronomy and Space Science, University of Science and Technology of China, Hefei, Anhui 230026, China }

\abstract {
We analyzed 12-year {\it Fermi} Large Area Telescope  $\gamma$-ray data in the inner Galaxy centered at ($l=30^{\circ}$, $b=0^{\circ}$) and ($l=330^{\circ}$, $b=0^{\circ}$). We found significant hardening of the spectrum of the diffuse \gray emission in these regions as previously reported. We further deduced that the diffuse $\gamma$ rays can be divided into two components from the likelihood analysis. One component is associated with the total gas column density and reveals a soft spectrum, while the other is associated with the \ion{H}{ii} gas and presents a hard spectrum.  Assuming the diffuse $\gamma$-ray emissions are mainly produced through the interaction between cosmic rays (CRs) and the ambient gas, these two components are produced by the CR populations with spectral indices of $2.8$ ("soft") and $2.3$ ("hard"), respectively.  We argue that the hard CR population may come from the vicinity of the CR accelerators. The soft CR population has a similar spectral shape and density as measured in the solar neighborhood, which implies a uniform CR "sea" with a similar density and spectral shape in the Galaxy. }
 \maketitle

\section{Introduction}

Cosmic rays (CRs) are believed to be well mixed in the magnetic field in our Galaxy and interact with the cold gas in the interstellar medium (ISM) to produce diffuse \gray emissions. Thanks to the progress in the \gray astronomy, such diffuse \gray emissions have been well studied in recent years \citep{fermi_diffuse, yang16}. Taking into account the gas content derived from molecular and atomic line spectrometry, one can obtain information on the CR distribution on the Galactic scale.  By assuming a cylindrically symmetric distribution,  \citet{yang16} and \citet{fermi_diffuse} found that  CRs are strongly inhomogeneous in our Galaxy. The radial distribution of CRs peaks at the galactocentric distance of about 4 kpc, and the derived CR spectra in the inner Galaxy are significantly harder. 

The spatial distribution of CRs has direct implications on the understanding of their origin and propagation. Generally speaking, the diffusion process will make the CR distribution smoother than the distribution of CR sources, but it cannot erase all hints of the source distributions \citep[see, e.g.,][]{strong98}. Thus, the sharp peak of the derived CR distributions requires a significant modification to the "standard" CR propagation model in our Galaxy. As an example, \citet{recchia16} invoked the CR-driven wind in the Galactic plane to explain the observed CR radial profile. 

However, whether such inhomogeneity is a global effect or caused by individual sources is still unclear. \citet{aharonian20} used the \gray observations of individual molecular clouds to derive the CR density therein, and unveiled a homogeneous “sea” of CRs with a constant density and spectral shape, especially for the galactocentric distances exceeding 8~kpc, as well as the Sagittarius B complex, in the region of the Galactic center (GC). However, in the region with galactocentric distances between 4 and 6 kpcs, the derived CR densities reveal significant deviation from the "sea" and each other. One possible explanation is that larger amounts of particle accelerators within this region cause the enhancement of CR density in their vicinity.  The distributions of potential CR accelerators, such as supernova remnants (SNRs), pulsars, and OB stars, do reveal strong galactocentric radius dependence \citep{green15,yusifov04,bronfman00}. Particularly, the distribution of OB stars presents a similar peak at about 4~kpc \citep{bronfman00}, which makes these massive star clusters an attractive source population for addressing the higher CR density and harder CR spectra in this region.  Indeed, young massive star clusters (YMCs) were recently identified as a new population of \gray sources and are believed to be an alternative source of the Galactic CRs \citep{aharonian19}.  In several such systems, including the Cygnus cocoon \citep{Ackermann11, aharonian19}, Westerlund 1 \citep{Abramowski12}, NGC 3603 \citep{Yang17}, Westerlund 2 \citep{Yang18} and 30 Dor C \citep{Abramowski15}, extended \gray emissions with hard spectra have been detected from GeV to TeV.  Besides this, most of the observed \grays reveal a good spatial correlation with the \ion{H}{ii} region, which is formed due to the photoionization of these massive stars  \citep{Yang17,aharonian19,sun20a,sun20b}.  Due to strong extinction in the Galactic plane, there may be numerous unknown YMCs exist in the Galaxy.  Thus, if there are additional contributions from the unknown YMCs, we expect the \gray emission to present similar spatial distribution as the \ion{H}{ii} gas.

To test such a hypothesis, we chose two regions centered at $(l,b)=(30^{\circ}, 0^{\circ})$ (region I) and $(l,b)=(330^{\circ}, 0^{\circ})$ (region II) to investigate the diffuse \gray emissions as well as the distribution of gases in different phases. We chose these regions because they are near the tangent point of the 4-kpc ring, and the aforementioned CR excess should have a strong impact on the total \gray fluxes in these directions. In the following, we give a detailed analysis of the 12-year \fermi data toward these regions and discuss the gas and CR content therein. The paper is organized as follows.  In Sect.\ref{sec:Gas}, we investigate the gas distribution in this region. In Sect.\ref{sec:gray}, we present details of the \gray data analysis. In Sect.\ref{sec:discussion}, we discuss the implications of our analysis results.

\section{Gas tracers}
\label{sec:Gas}

The 21 cm \ion{H}{i} line and 2.6 mm CO line are widely used tracers for atomic hydrogen and molecular hydrogen, respectively.  In this work, we used the data from  the Galactic CO survey of \citet{dame01} with the 
CfA 1.2m-millimeter-wave Telescope,  and the HI4PI Survey on HI gas \citep{HI4PI16}. 
For the CO  data, we used the standard assumption of a linear relationship between the velocity-integrated  CO intensity, $W_{\rm CO}$, and the column density of molecular hydrogen, N(H$_{2}$). 
The conversion factor $X_{\rm CO}$ may be different in different parts of the Galaxy; therefore, we used the values derived from \citet{fermi_diffuse} for different galactocentric distances, such as $X_{\rm CO} = 0.5 \times 10^{20}~\rm cm^{-2}(K~km/s)^{-1} $ below 1.5 kpc and  $X_{\rm CO} = 1.5 \times 10^{20}~\rm cm^{-2}(K~km/s)^{-1} $ above. 

For the HI data, we use the following equation: 
\begin{equation} 
N_{HI}(v,T_s)=-log \left(1-\frac{T_B}{T_s-T_{bg}}\right)T_sC_i\Delta v \ ,
\end{equation}
where $T_{bg}\approx2.66$~K is the brightness temperature of the cosmic microwave background radiation at 21cm, and
$C_i = 1.83 \times 10^{18} \rm  cm^{2}$.  In the case  when $T_B > T_s-5~\rm K$, 
we truncated $T_B$ to $T_s-5~\rm K$; 
$T_s$ was chosen to be 150 K. The systematic uncertainties due to the different spin temperatures 
are discussed in \citet{fermi_diffuse_old} and \citet{fermi_diffuse}. The effect, however, is  quite small in most regions of the sky. 

In order to investigate the galactocentric radial distribution of CRs, we need to divide the gas distribution into different rings around the GC.  For this purpose,  we used the following  relation: 
\begin{equation}
V_{LSR}=R_{\odot}( \frac{V(R)}{R}-\frac{V_{\odot}}{R_{\odot}})sin(l)cos(b),
\label{eq:rot}
\end{equation}
where $V_{LSR}$ is the radial velocity  with respect to the local standard of rest,  $R$ is the galactocentric distance, $V(R)$ is the Galactic rotational curve,  and $l$ and $b$  are the galactic coordinates in the line  of sight (LOS). We adopted the rotational curve parameterized in \citet{clemens85}; $V_{\odot}$ and $R_{\odot}$ are fixed to 220 km/s and 8.5 kpc, respectively. By applying this to both CO and HI data, we can transform the velocity information into the galactocentric distance of the gas. According to Fig. 8 of \citet{fermi_diffuse}, we split the gas into galactocentric distance bins; for example, below and above 4.5 kpc in the analysis below. It should be noted that  Eq.\ref{eq:rot}  
allows  emission  from  the forbidden velocity zones in the CO and HI data. 
The reason could be  the non-circular motion of the gas.  This component contains only a small fraction of total gas; therefore, for simplicity, we assigned it to the local rings.

\begin{figure*}[ht]
\centering
\includegraphics[width=0.45\textwidth]{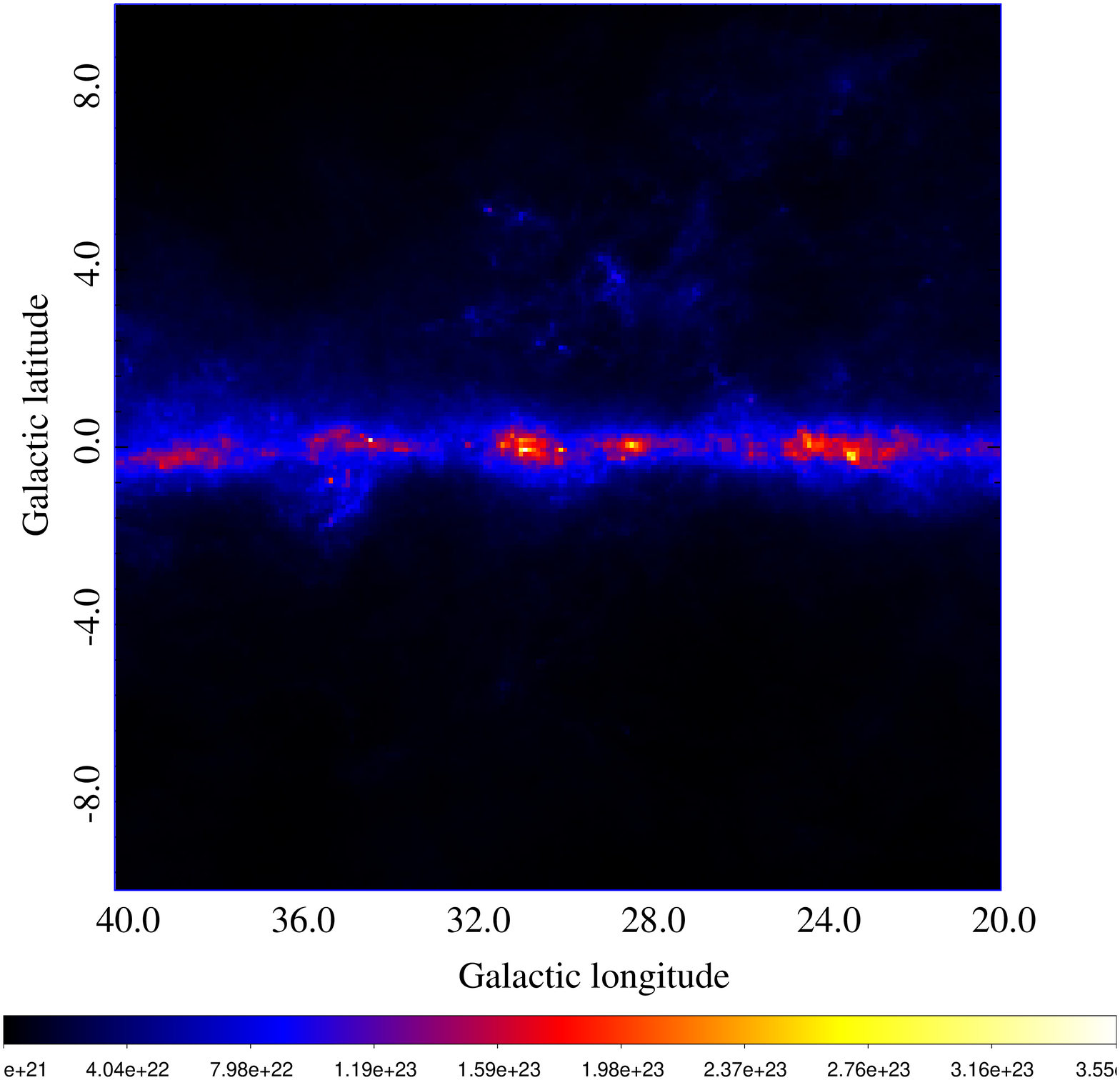}
\includegraphics[width=0.45\textwidth]{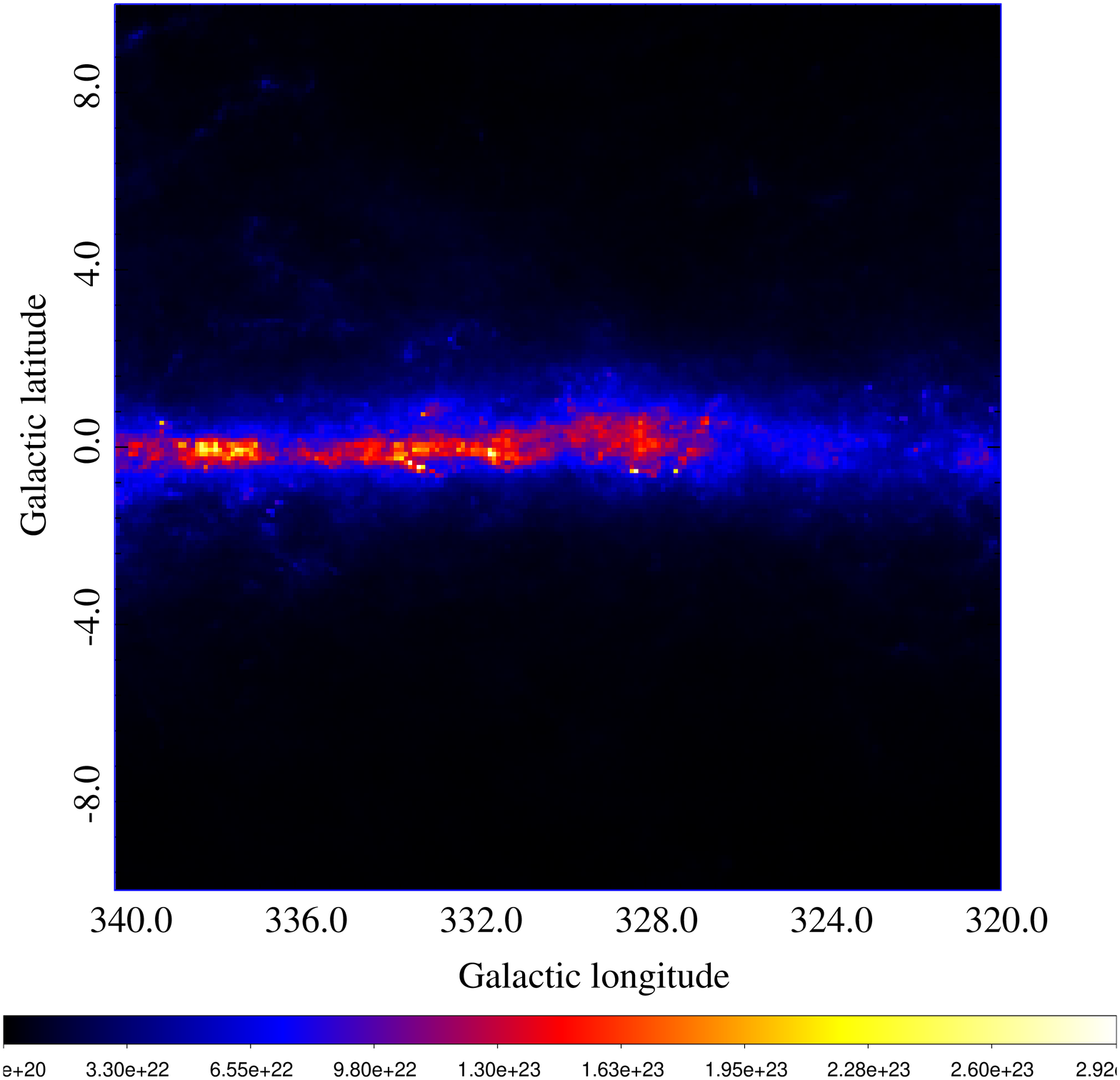}
\includegraphics[width=0.45\textwidth]{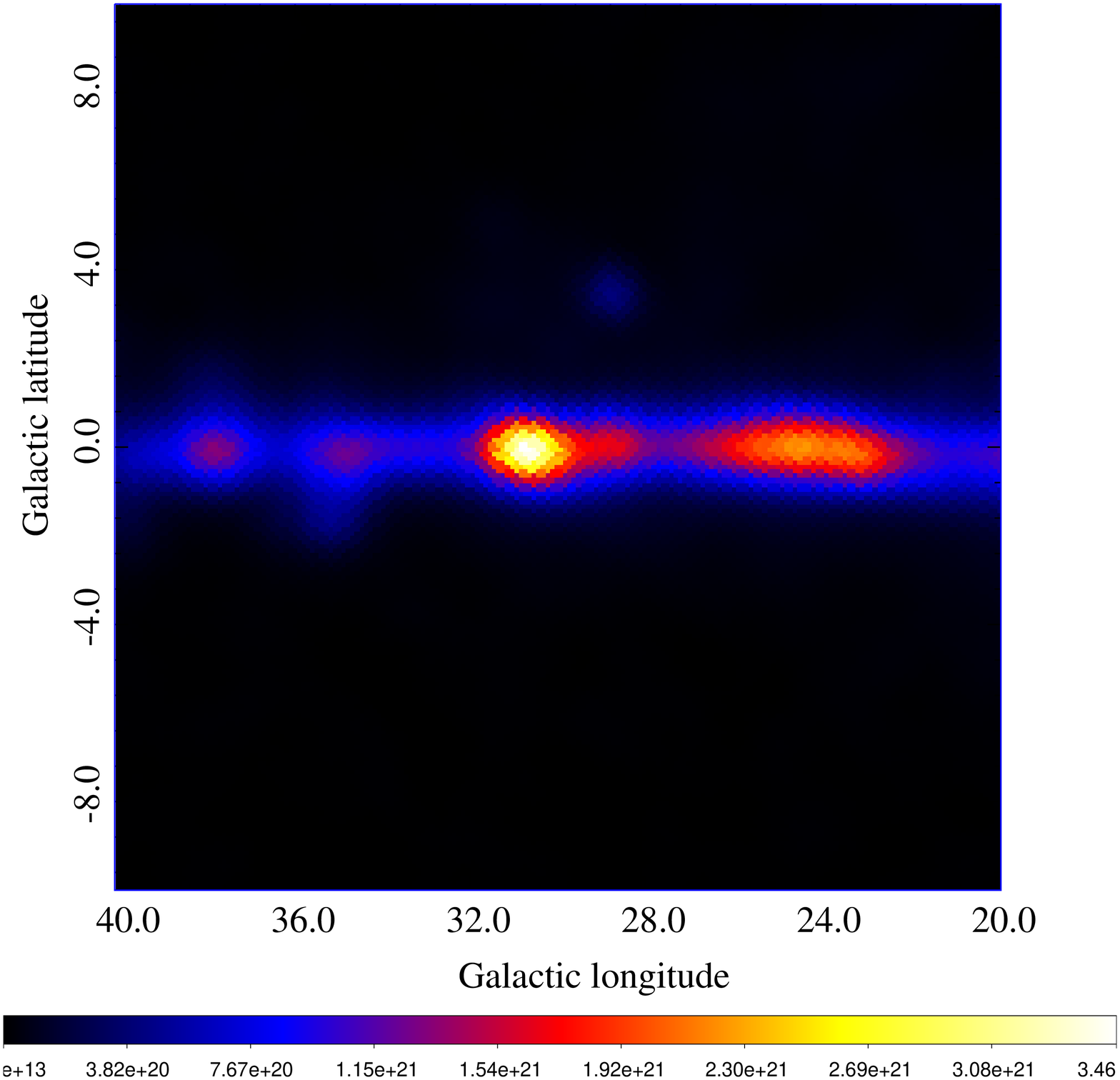}
\includegraphics[width=0.45\textwidth]{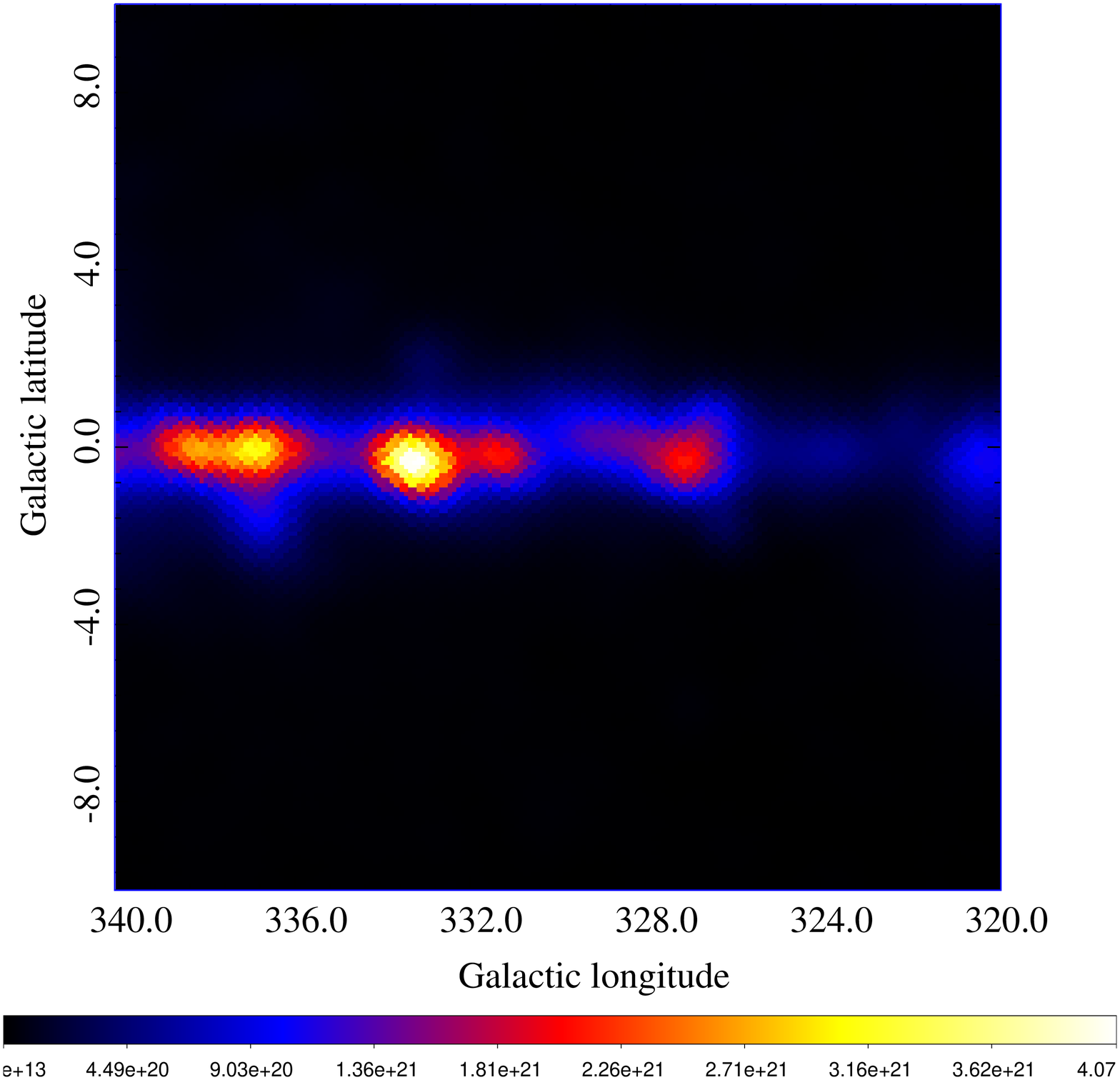}
\caption {Dust and \ion{H}{ii} templates used in \fermi data analysis.
Top panel: Derived total gas column density maps  (in units of ${\rm cm}^{-2}$)  from Planck dust opacity for region I  (left) and region II (right).  Bottom panel: Derived \ion{H}{ii} gas density maps  (in units of ${\rm cm}^{-2}$) from the joint analysis of {\it Planck}, {\it WMAP}, and 408 MHz observations for region I  (left) and region II (right).
}
\label{fig:dust}
\end{figure*}

For different reasons, the neural gas cannot always be traced by CO and \ion{H}{i} observations \citep{grenier05}.  In such cases (e.g. in optically thick clouds), the infrared emission from cold interstellar dust provides an alternative and independent measurement of the gas column density. In the analysis below, we also used the dust column map as gas template for cross-checking. 
According to Eq.~(4) of \cite{planck11}, the relation between the dust opacity and the column density 
can be approximated as
\begin{equation}\label{eq:dust}
\tau_{\rm M}(\lambda) = \left(\frac{\tau_{\rm D}(\lambda)}{N_{\rm H}}\right)^{\rm dust}[N_{{\rm H I}}+2X_{\rm CO}W_{\rm CO}],
 \end{equation}
where $\tau_{\rm M}$ is the dust opacity as a function of the wavelength  $\lambda$,  $(\tau_{\rm D}/N_{\rm H})^{\rm dust}$ is the reference dust emissivity measured in low-$N_{\rm H}$ regions, $W_{\rm CO}$ is the integrated brightness temperature of the CO emission, and $X_{\rm CO}=N_{\rm H_2}/W_{\rm CO}$ is the  $\rm {H_2/CO}$ conversion factor.
The substitution of  the latter into Eq.~(\ref{eq:dust})  gives 
\begin{equation}
N_{\rm H} =N_{\rm HI}+2 N_{\rm H_2} = \tau_{\rm M}(\lambda)\left[\left(\frac{\tau_{\rm D}(\lambda)}{N_{\rm H}}\right)^{\rm dust}\right]^{-1}. 
\end{equation}
Here  for the dust emissivity at $353~\rm GHz$,   we used  $(\tau_{\rm D}/N_{\rm H})^{\rm dust}_{353{\rm~GHz}}=(1.18\pm0.17)\times10^{-26}$~cm$^2$  taken from Table~3 of \cite{planck11}. 
The derived total gas column density distributions in region I and region II are shown in the top panel of Fig.\ref{fig:dust}.

We also used the dust data to derive the residual templates for the missing "dark" gas component. We fit the dust opacity maps as a linear combination of HI and CO maps, and, in this way, find  the residual map to be the dark gas template. Then, we iterated this fit by including the dark gas template until  convergence is achieved \citep{yang16}. 
This method is similar to the derivation of  the $E(B-V)_{res}$ templates used by the Fermi-LAT collaboration \citep{fermi_diffuse}, 
where instead of the extinction maps we used the dust opacity maps.   Indeed, $E(B-V)$ has a nearly perfect linear correlation with the dust opacity, especially in higher column regions \citep{planck13-11}. Thus, our method and the method used in \citet{fermi_diffuse} should give similar results. 

We used the free-free emission map obtained from the joint analysis of {\it Planck}, {\it WMAP}, and 408 MHz observations \citep{Planck16} to derive the map of \ion{H}{ii} column density.
First, we converted the emission measure into free-free intensity ($I_{\nu}$) by using the conversion factor at 353-GHz in Table 1 of \citet{Finkbeiner03}.
Then, we applied Eq.~(5) from \citet{Sodroski97},
\begin{equation}
\begin{aligned}
N_{\ion{H}{ii}} = &1.2 \times 10^{15}\ {\rm cm^{-2}} \left(\frac{T_{\rm e}}{1\ \rm K}\right)^{0.35} \left(\frac{\nu}{1\ \rm GHz}\right)^{0.1}\left(\frac{n_{\rm e}}{1\ \rm cm^{-3}}\right)^{-1} \\
&\times \frac{I_{\nu}}{1\ \rm Jy\ sr^{-1}},
\end{aligned}
\end{equation}
to convert the free-free intensity into column density in each pixel, with a frequency at $\nu = \rm 353\ GHz$, and an electron temperature of $T_{e} =\rm 8000\ K$. 
This equation also shows that the \ion{H}{ii} column density is inversely proportional to the effective density of electrons $n_{\rm e}$. Here, we chose $n_{\rm e}=2\ \rm cm^{-3}$ \citep{Sodroski97} to estimate the upper limit of the \ion{H}{ii} column density. We note that $n_{\rm e}$ varies in different positions of the Galaxy; however, in this study, most of the \ion{H}{ii} column densities are from the inner Galaxy. As discussed in \citet{Sodroski97}, the assumption of using $n_{\rm e}=2\ \rm cm^{-3}$ as the effective electron density in the \ion{H}{ii} regions in the inner Galaxy produce the same  gas-to-dust mass ratio for the \ion{H}{ii} component  as the mean values of the gas-to-dust mass ratio for the H2 and HI components within the same ranges of radial distance. Thus, in the following study, we fixed $n_{\rm e}=2\ \rm cm^{-3}$ to derive the column density of the \ion{H}{ii} component. The derived \ion{H}{ii} column density maps of both regions are  shown in the bottom panel of Fig.~\ref{fig:dust}.

\begin{figure*}[ht]
\centering
\includegraphics[width=0.45\textwidth]{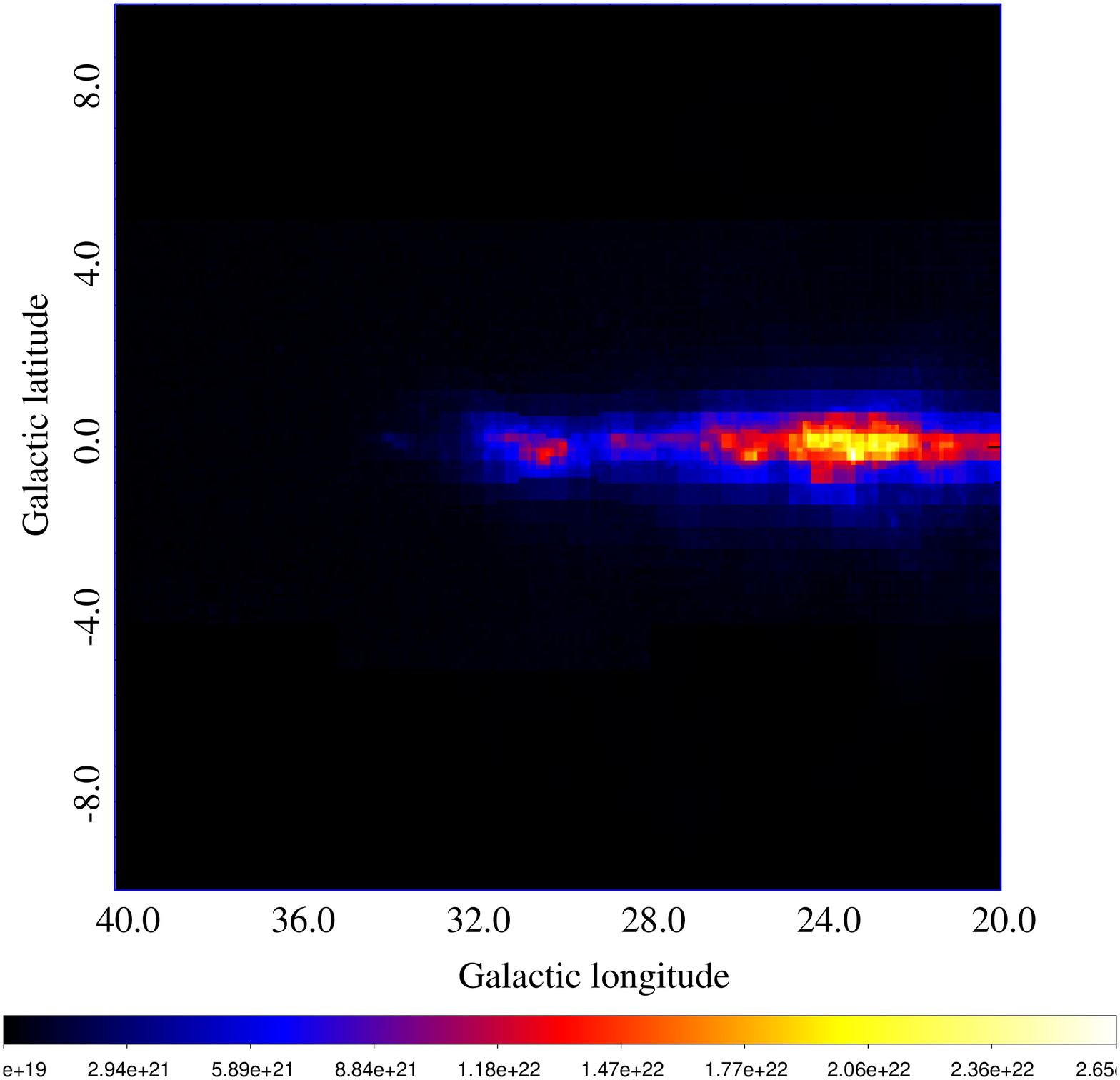}
\includegraphics[width=0.45\textwidth]{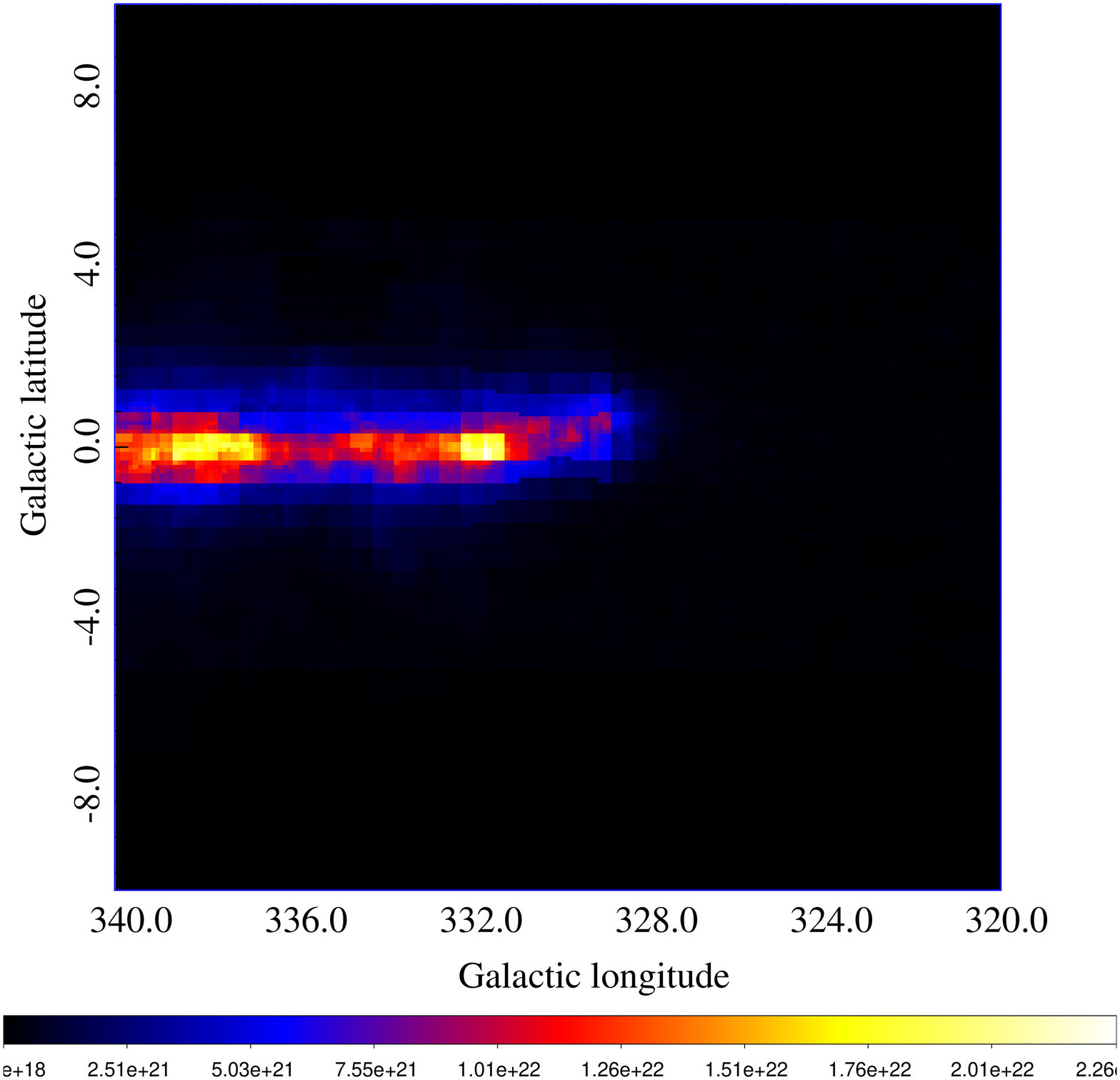}
\includegraphics[width=0.45\textwidth]{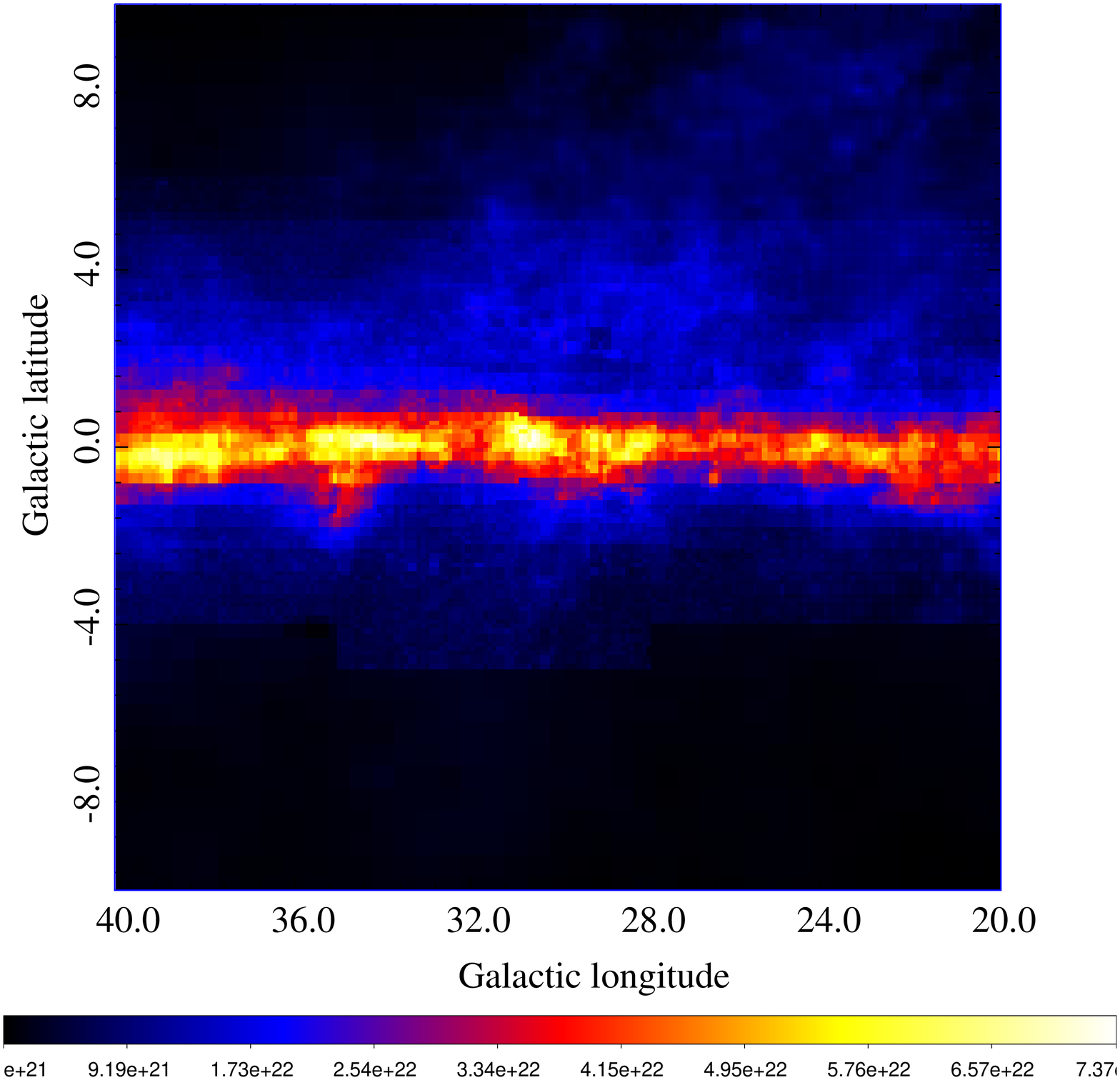}
\includegraphics[width=0.45\textwidth]{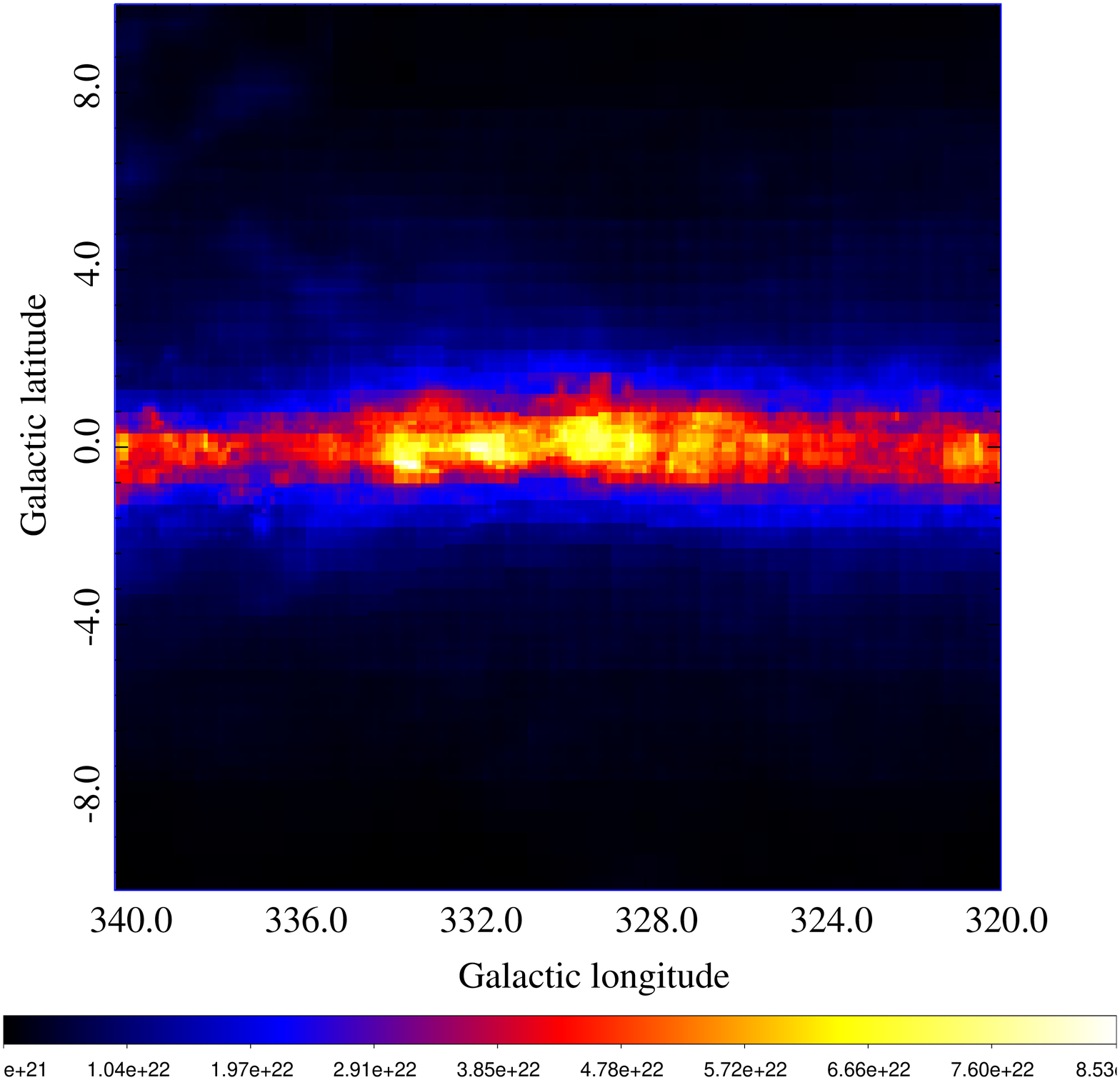}
\caption {Gas column density maps (in units of ${\rm cm}^{-2}$)  below 4.5\,kpc (top panel) and  above 4.5\,kpc (bottom panel) for region I (left) and region II (right), which are derived from CO and HI observation data \citep{dame01,HI4PI16} using method described in Sect.\ref{sec:Gas}.}
\label{fig:gas}
\end{figure*}

\begin{figure*}[ht]
\centering
\includegraphics[width=0.45\textwidth]{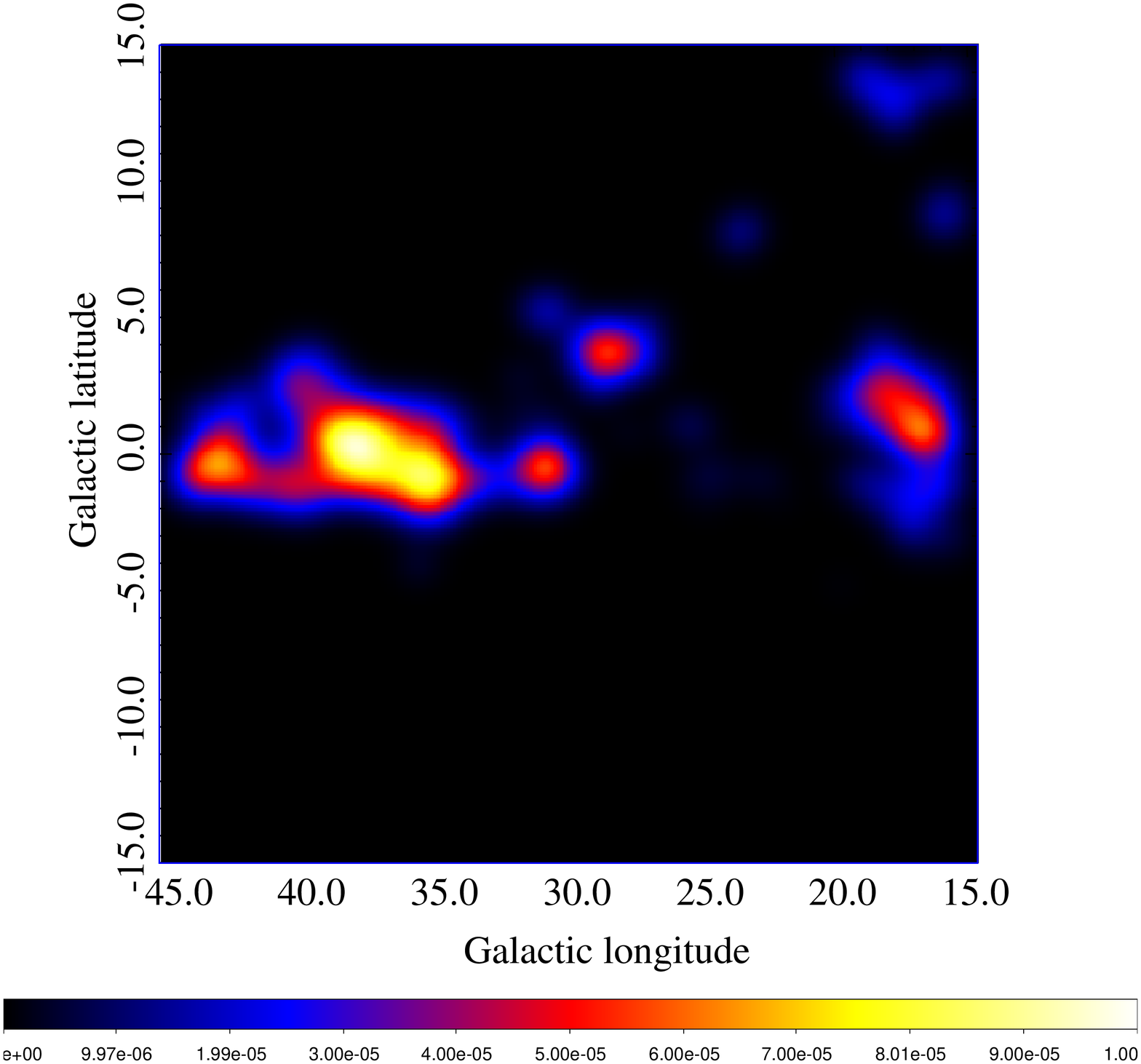}
\includegraphics[width=0.45\textwidth]{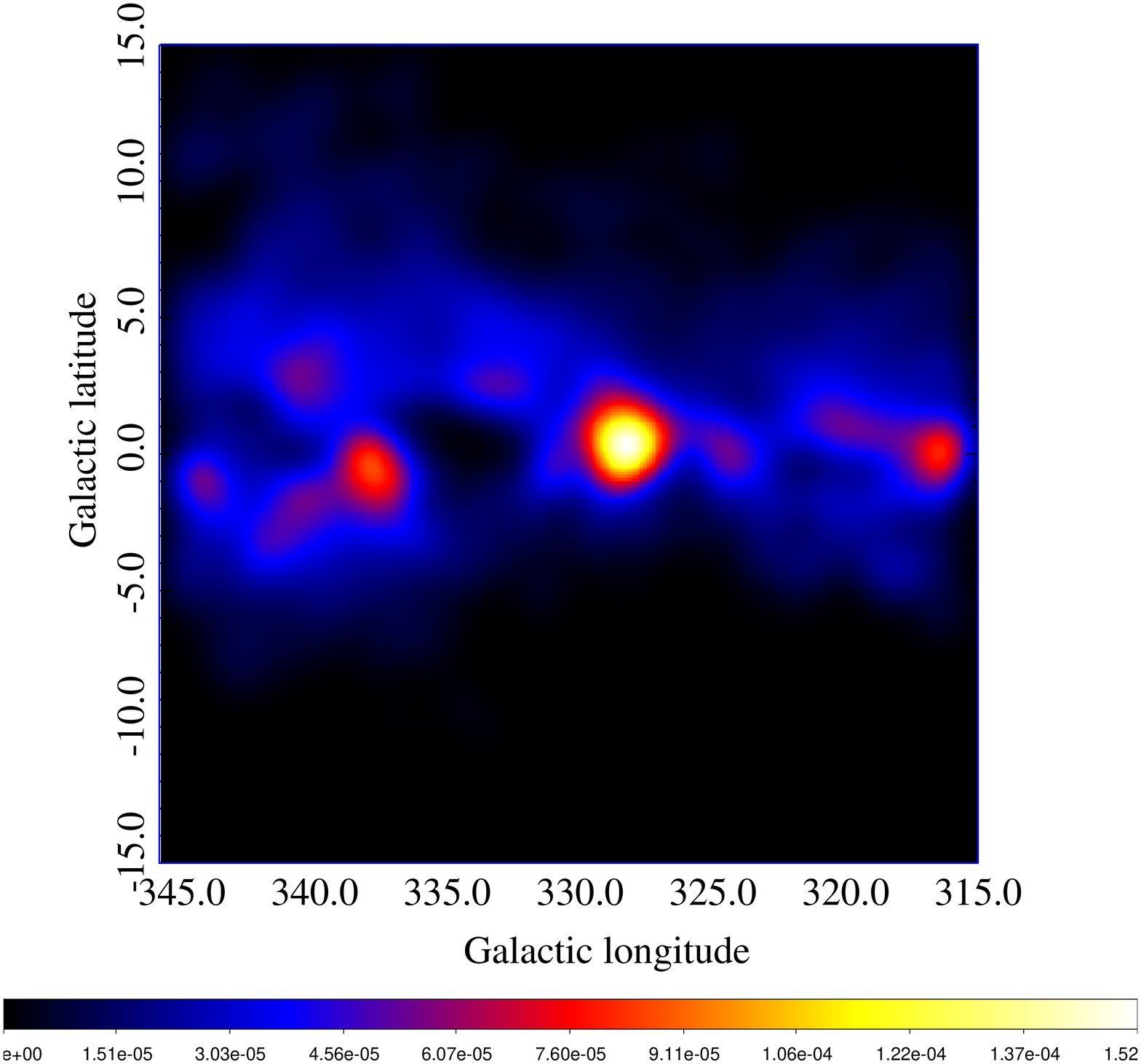}
\caption { Dark gas templates for region I (left) and region II (right), which are derived from  dust data using the method described in Sect.\ref{sec:Gas}.}
\label{fig:dnm}
\end{figure*}

\section{\fermi data analysis}
\label{sec:gray}
We collected the \fermi Pass 8 data from August 4, 2008 (MET 239557417) until  October 28, 2020 (MET 625621368), and used the Fermitools from the Conda distribution\footnote{https://github.com/fermi-lat/Fermitools-conda/} together with the latest version of the instrument response functions (IRFs) {\it P8R3\_SOURCE\_V3} for the \gray emission analysis.
Two $20\deg \times 20\deg$  square regions centered at the position of  ($l=30^{\circ}$, $b=0^{\circ}$) and ($l=330^{\circ}$, $b=0^{\circ}$) were chosen as two regions of interest (ROI). 
Here, we selected the "source" class events, then applied the recommended data cut expression $\rm (DATA\_QUAL > 0) \&\& (LAT\_CONFIG == 1)$ to exclude time periods when some spacecraft events affected the data quality. Moreover, to reduce the background contamination from the Earth's albedo, only the events with zenith angles under $90\deg$\ were included in the analysis.
\begin{figure*}[ht]
\centering
\includegraphics[width=0.45\textwidth]{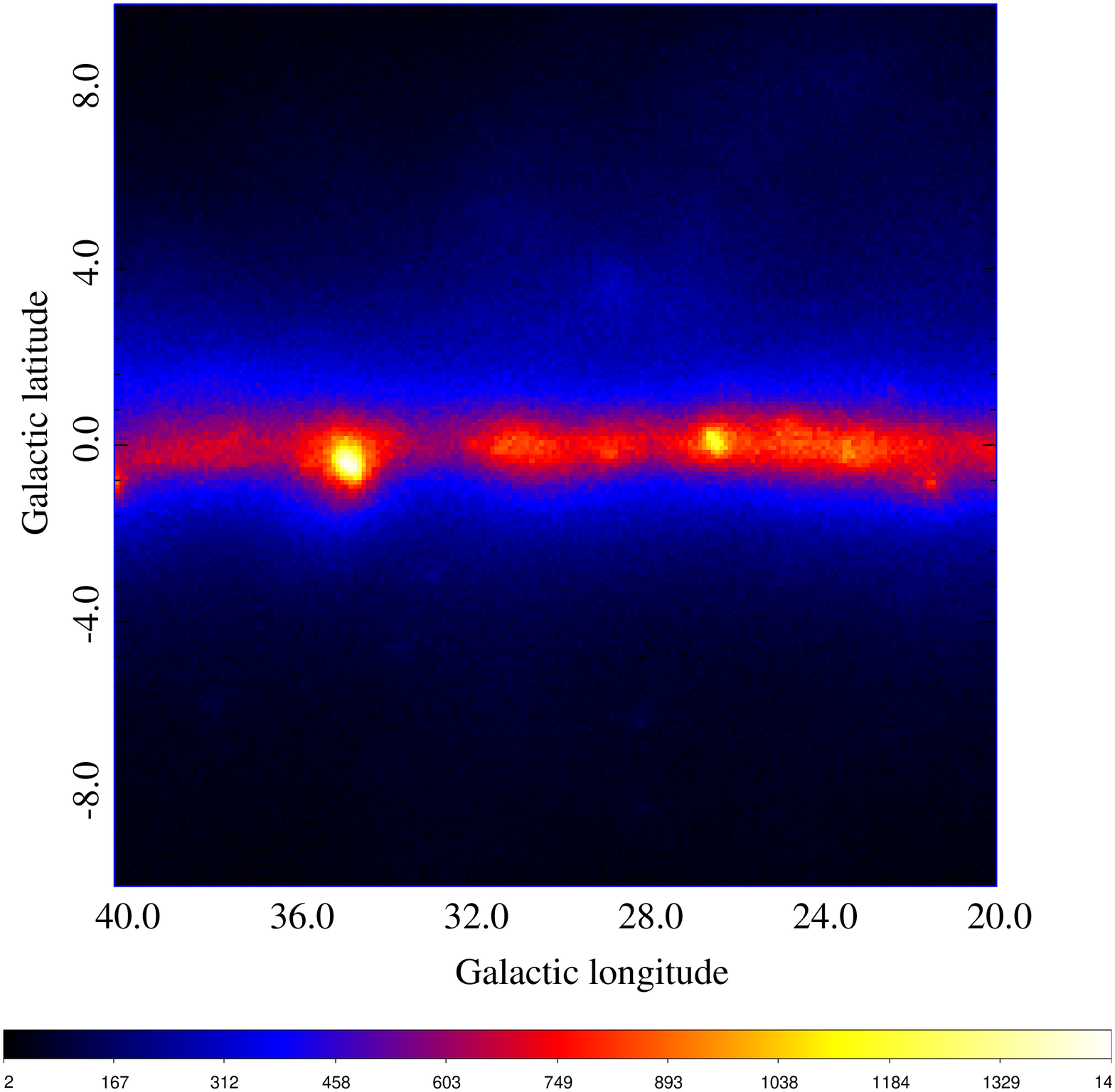}
\includegraphics[width=0.45\textwidth]{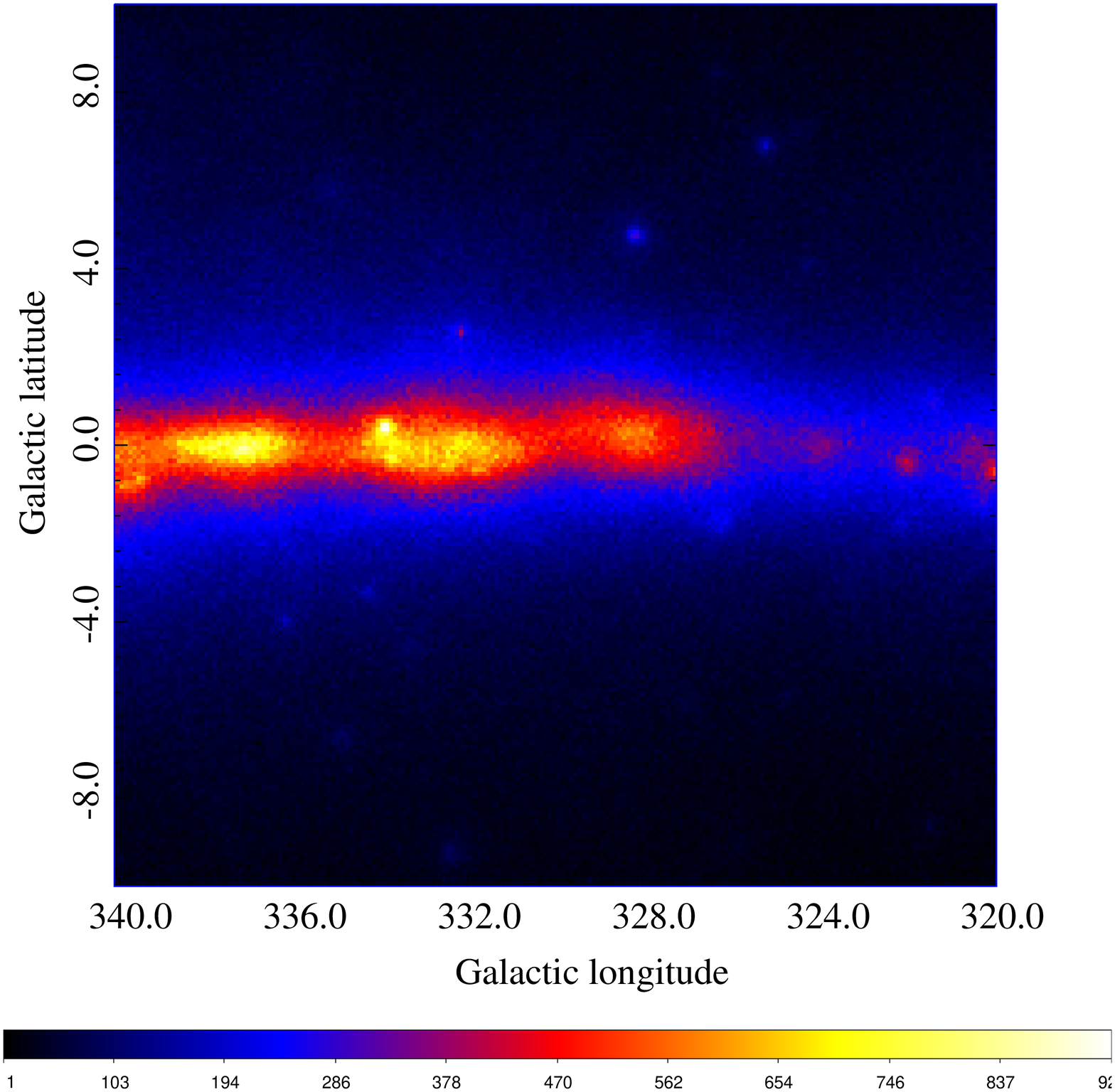}
\caption { 
Observed counts map of \grays in the energy range of 0.2--200\,GeV for region I  (left) and region II (right). 
}
\label{fig:cmap}
\end{figure*}

For each region, the  background model includes the sources in the \fermi\ ten-year Source Catalog  \citep[4FGL-DR2,][]{4fgl_dr2} within the ROI enlarged by 10\deg, and the normalizations  and spectral parameters were left free for sources with significance $>5\sigma$ within the ROI.

\subsection{Diffuse \gray emission}
As for the diffuse background components, we first used  the \fermi Galactic diffuse background model ({\it gll\_iem\_v07.fits}), hereafter referred to as the "Fermi" model.
Since we are interested in the diffuse \gray emissions,  we also built our own Galactic diffuse background models for cross-checking.  The diffuse \gray emission mainly comes from the pion decay process induced by the inelastic collision between CR protons and ambient gas,  as well as the inverse Compton (IC) scattering of CR electrons 
in the interstellar radiation fields (ISRFs). We calculated the IC component using GALPROP\footnote{\url{http://galprop.stanford.edu/webrun/}} \citep{galprop}, which uses information regarding CR electrons and ISRFs. For the GALPROP webrun we used the parameter set $^SY ^Z4^R30^T150^C2$ in \citet{fermi_diffuse_old} as a fiducial model. We also used the 128 parameter sets in \citet{fermi_diffuse_old} to generate the corresponding IC emission templates;  and for each region, we chose the best template  by comparing the maximum likelihood achieved by each likelihood fitting process.

For the pion decay component, as a zero-order approximation, the CRs are homogeneously distributed in the CR sea in our Galaxy. Thus, we can use the gas distribution map to model the pion decay component of the diffuse \gray emission. As mentioned above, to avoid the uncertainties of dark gas and \ion{H}{i} spin temperatures, we used the gas column density maps derived from the Planck dust opacity. 
Instead of the  template {\it gll\_iem\_v07.fits} in the Fermi models, models using total gas distribution maps derived from Planck dust opacity ( as shown in the top panels of Fig.~\ref{fig:dust}), and IC emission templates are referred to as "dust" models.

However, since the metallicity varies with galactocentric distance, the dust-to-gas ratio may also vary in the Galaxy. Furthermore,  as derived in \citet{fermi_diffuse}, the spectra and density of CRs vary in galactocentric distance. According to the results in \citet{fermi_diffuse}, both the spectra and density of CRs and the $X_{CO}$ factor vary significantly below and above 4.5 kpc. Thus, we split the gas into to rings with galactocentric radius below and above 4.5 kpc and include both of them in the pion decay component, and the gas column density distributions within different galactocentric distances are shown in Fig.\ref{fig:gas}.  In this model, as mentioned above, a dark gas component must be invoked to account for the gas that cannot be traced by CO and $21~\rm cm$ observations. Thus, we also added a dark gas component as described in the last section. 
Models that replace the total gas distribution map with split gas distribution maps and the dark gas component maps (as shown in Fig.\ref{fig:dnm}) are referred to as "gas" models. 
Finally, as we investigated the possible contribution of an \ion{H}{ii} component in the diffuse \gray emissions, we also added the \ion{H}{ii} column density map (as shown in the bottom panel of Fig.\ref{fig:dust}) as a diffuse template in the following analysis. 
In the following analysis, we used a log-parabola function ($dN/dE=N_0 (E/E_{\rm b})^{-\alpha -\beta log(E/E_{\rm b})}$) as the energy spectrum of the gas template and left both the normalization and spectral parameters free in the likelihood fitting below.

\subsection{Results}

Using photons of energies within 0.2--200~GeV, we performed the standard likelihood analysis of \fermi data with the  4FGL-DR2 sources and diffuse components mentioned above.
To check the possible contribution from the \ion{H}{ii} component, we performed the analysis with and without a template derived from \ion{H}{ii} column density, respectively.
To compare the goodness of fit in these two models, we also calculated the Akaike information criterion (AIC) value for each model. AIC was defined as $AIC = -2 \rm log(likelihood) +2 k$, where k is the number of free parameters in the model. The log(likelihood) and corresponding AIC value for each fit are listed in Table.\ref{tab:logl}. In addition, we compared the overall maximum likelihood of the model with \ion{H}{ii} template (alternative hypothesis, $L$ ) and that of the model without  (null hypothesis, $L_{0}$)  to obtain the test statistics (TS) of the {\ion{H}{ii}} component which is defined as $-2{\rm log}(L_{0}/L)$ following \citet{Lande12}. Then, the corresponding  significance ($\sigma_{\ion{H}{ii}}$) of the \ion{H}{ii} component can be estimated as $\sim\sqrt{\rm TS_{\ion{H}{ii}}}$. The TS value and the significance of the additional {\ion{H}{ii}} component for each model are also listed in Table.\ref{tab:logl}.   

We found that in both regions the inclusion of the \ion{H}{ii} templates for dust and gas models improves the fit dramatically, and the results are even improved for the Fermi models.  However, the overall results reveal that the gas models cannot fit the data as well as the Fermi or dust models. Furthermore, by subtracting the fit model map calculated using {\it gtmodel} from the observed \gray counts map (as shown in Fig.\ref{fig:cmap}), we generated the residual maps of these different models. Then, we divided the residual counts  by the square root of the observed \gray counts in each pixel, which should be equal to the significance (in $\sigma$) of each pixel approximately. We used these residual significance (signal-to-noise, S/N) maps to compare the fitting quality of these models. The improvement can be also seen from the comparisons of the results with and without the \ion{H}{ii} templates, which are shown in Fig.\ref{fig:ts_reg1} and Fig.\ref{fig:ts_reg2}.  In addition, we used a Gaussian function  to fit the distribution of the S/N ratios of each model to check whether the model is a statistically good description of the data. As shown in the bottom panel of  Fig.\ref{fig:ts_reg1} and Fig.\ref{fig:ts_reg2}, each  distribution can be well described by the Gaussian function with a mean of $\sim$0.0 and standard deviation of $\sim$1.0, which indicates an unbiased significance distribution of the background.

\begin{table*}[htbp] 
\caption{Fitting results for different models of region I and region II} 
\label{tab:logl} 
\centering
\begin{tabular}{clccccc}
\hline
\hline
 Region& Model &~free parameters (k) &~log(likelihood)&~AIC & ${\rm TS}_{\rm \ion{H}{ii}}$&$\sigma_{\ion{H}{ii}}$\\
\hline
 \multirow{6}{*}{ I} &  Fermi&~506 &~15876971.5&~-31752931.0&~ &\\
 &Fermi+\ion{H}{ii} &~509&~15877085.4&~-31753152.8&~227.8 &15.1\\
 & dust &~509 &~15873005.8&~-31744993.6&~ &\\
&dust+\ion{H}{ii} &~512 &~15876432.2&~-31751840.4&~6852.8 &82.8\\
& gas&~515  &~15870347.9 &~-31739665.8&~ &\\
&gas+\ion{H}{ii} &~518 &~15872922.1&~-31744808.2&~5148.4 &71.8\\
\hline
 \multirow{6}{*}{II}& Fermi&~592 &~8009736.8&~-16018289.6&~ &\\
&Fermi+\ion{H}{ii}  &~595 &~8009738.5&~-16018287.0&~3.4 &1.8\\
 &dust &~595 &~8006053.7&~-16010917.4&~ &\\
&dust+\ion{H}{ii} &~598 &~8008499.1&~-16015802.2&~4890.8 &69.9\\
 &gas&~601 &~8005227.7&~-16009253.4&~ &\\
&gas+\ion{H}{ii}  &~604 &~8006935.4&~-16012662.8&~3415.4 &58.4\\
\hline
\end{tabular}
\end{table*}

\begin{figure*}[ht]
\centering
\includegraphics[width=0.45\hsize]{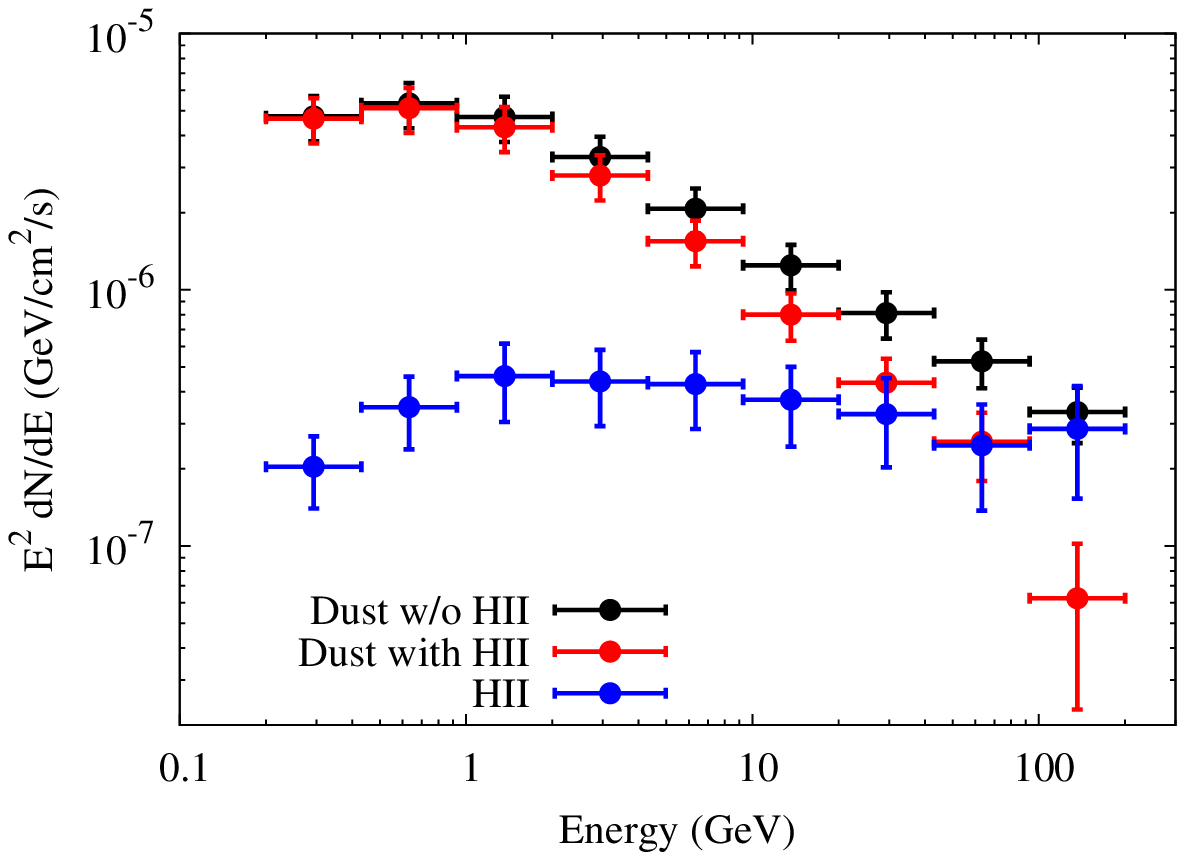}
\includegraphics[width=0.45\hsize]{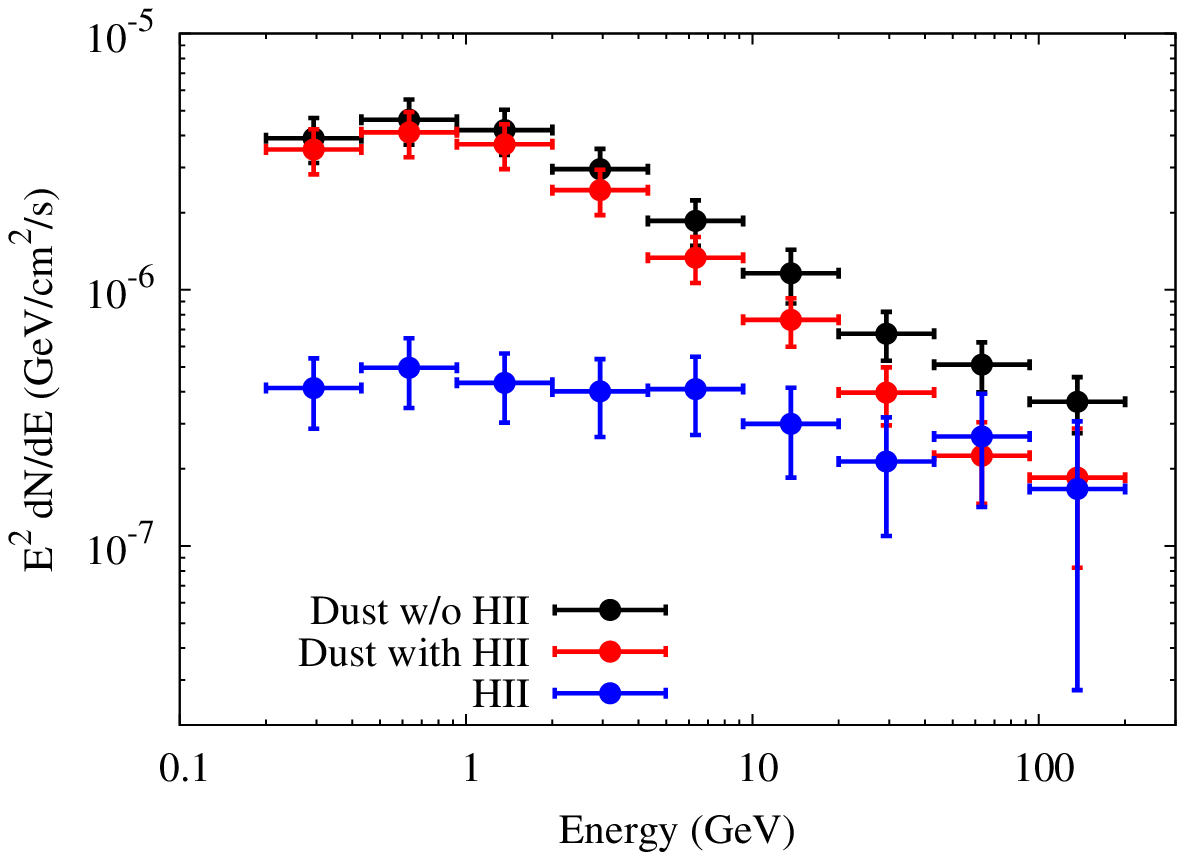}
\caption {SEDs of diffuse \gray emissions associated with different components for region I (left panel) and region II (right panel). The black and red points show the \grays associated with total gas (the dust template) for both dust models and dust+\ion{H}{ii} models, respectively.  The blue points represent the \grays related to \ion{H}{ii} gas (the \ion{H}{ii} template) obtained for dust+\ion{H}{ii} models.} 
\label{fig:spedust}
\end{figure*}

\begin{figure*}[ht]
\centering
\includegraphics[width=0.45\hsize]{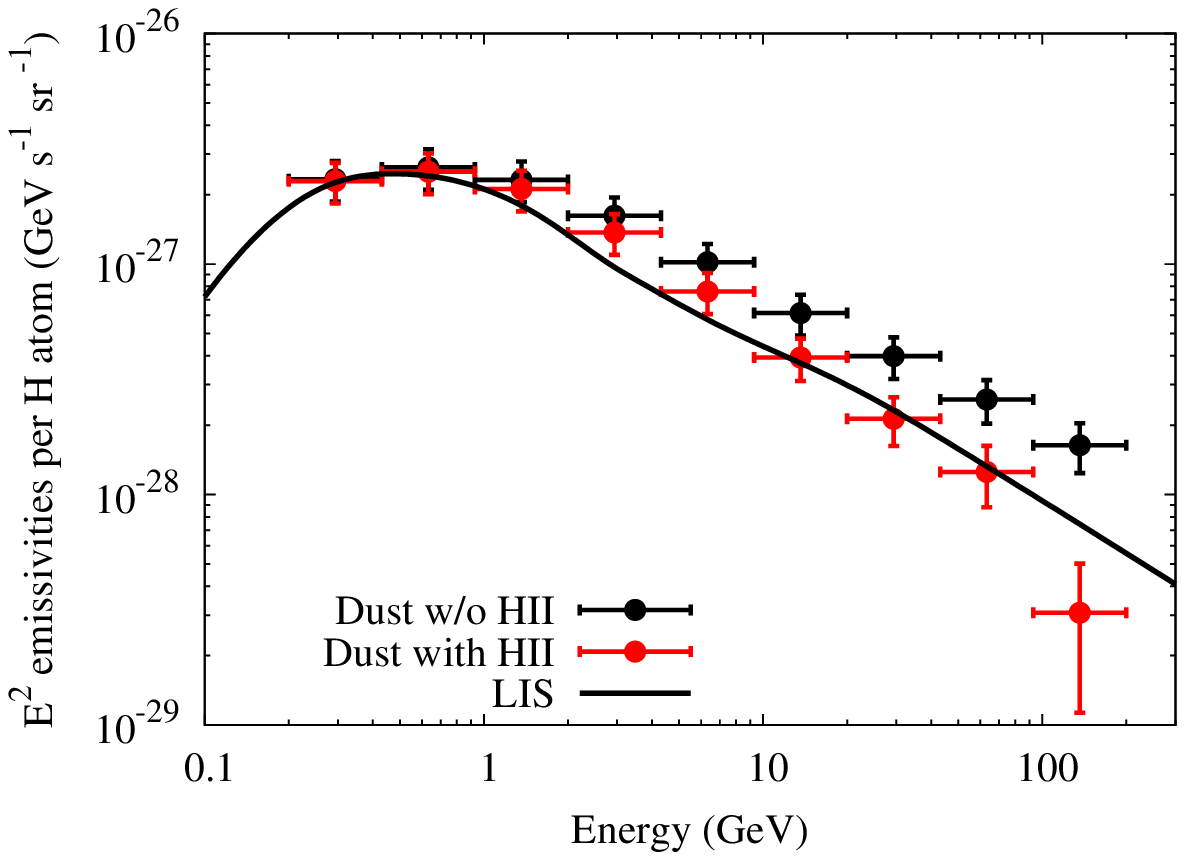}
\includegraphics[width=0.45\hsize]{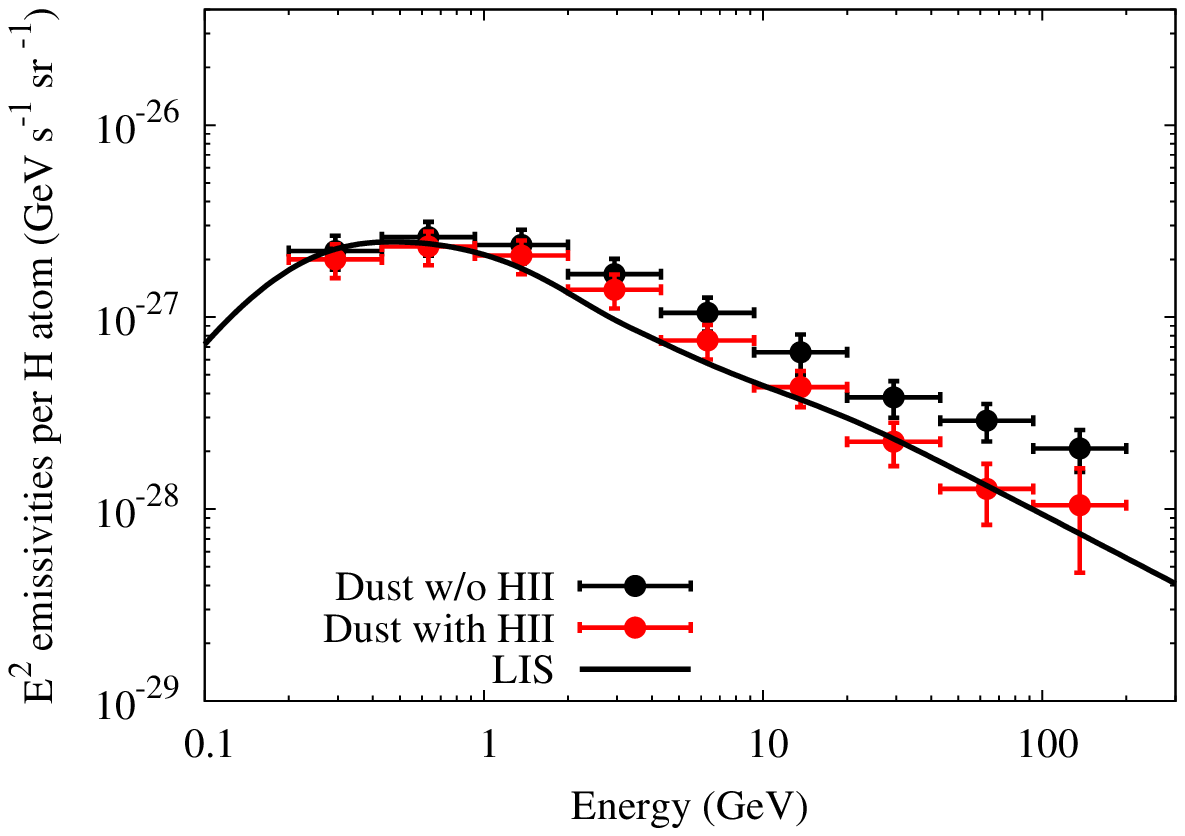}
\caption {Same as Fig.\ref{fig:spedust}, but the SEDs have been normalized to the \gray emissivity per H atom, which is proportional to the CR density. Also plotted as black curves are the predicted \gray emissivities for the local CR spectra (LIS). }
\label{fig:seddust}
\end{figure*}

\subsection{Spectral analysis}
\begin{figure*}[ht]
\centering
\includegraphics[width=0.45\hsize]{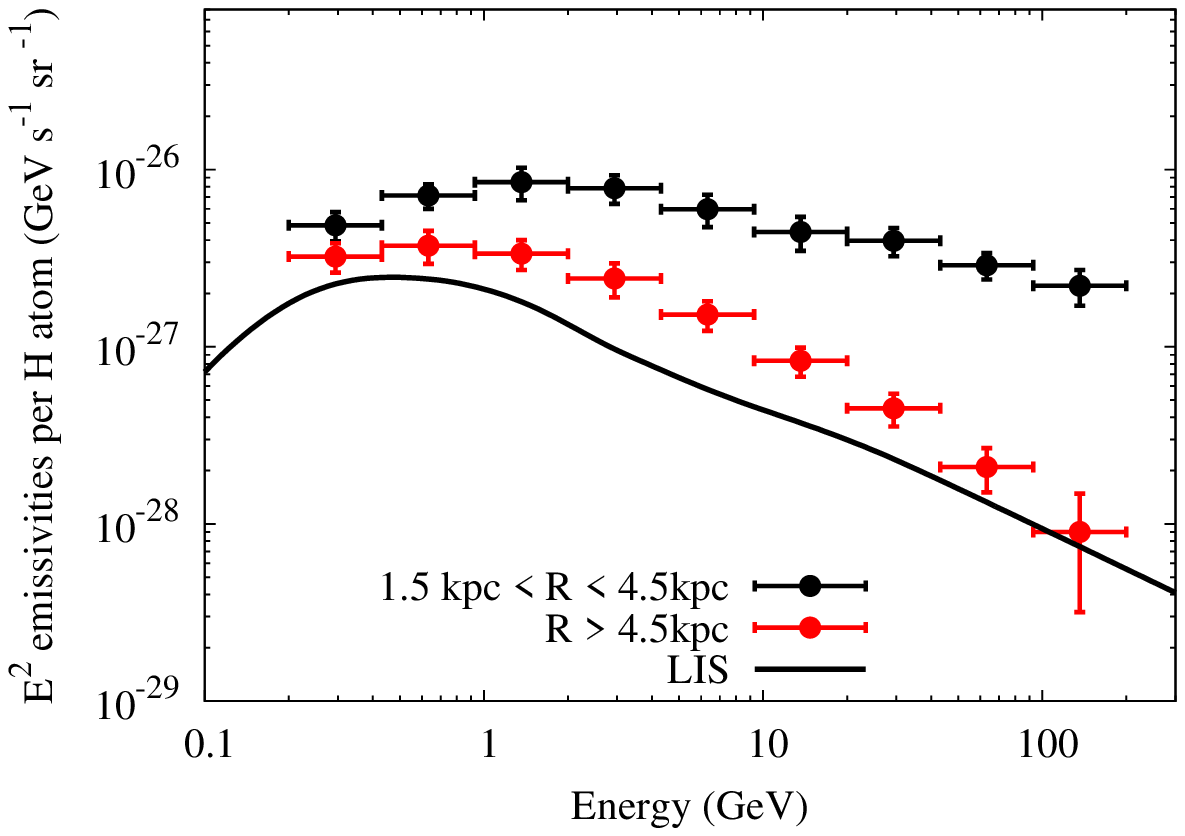}
\includegraphics[width=0.45\hsize]{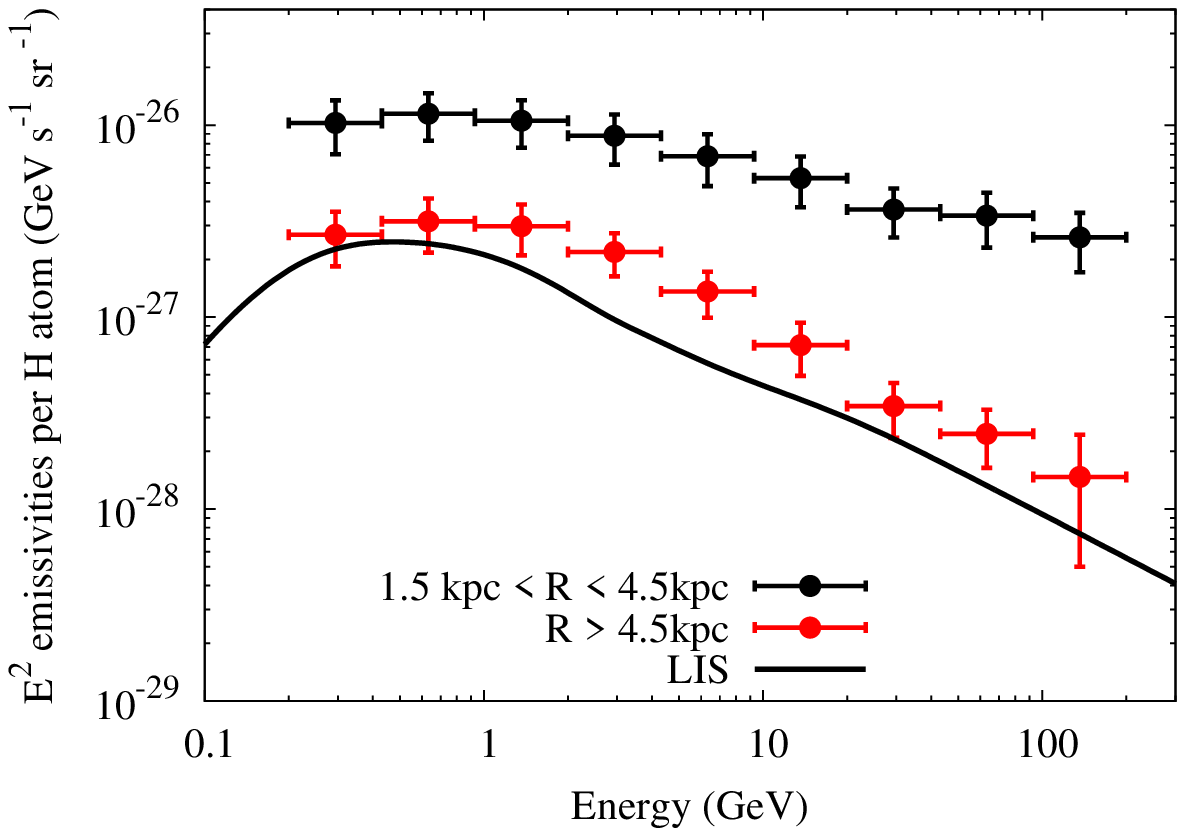}
\caption {\gray emissivities per H atom of diffuse \gray emission associated with the two gas rings in gas models for region I (left panel) and region II (right panel), respectively.}
\label{fig:sedgas}
\end{figure*}

\begin{figure*}[ht]
\centering
\includegraphics[width=0.45\hsize]{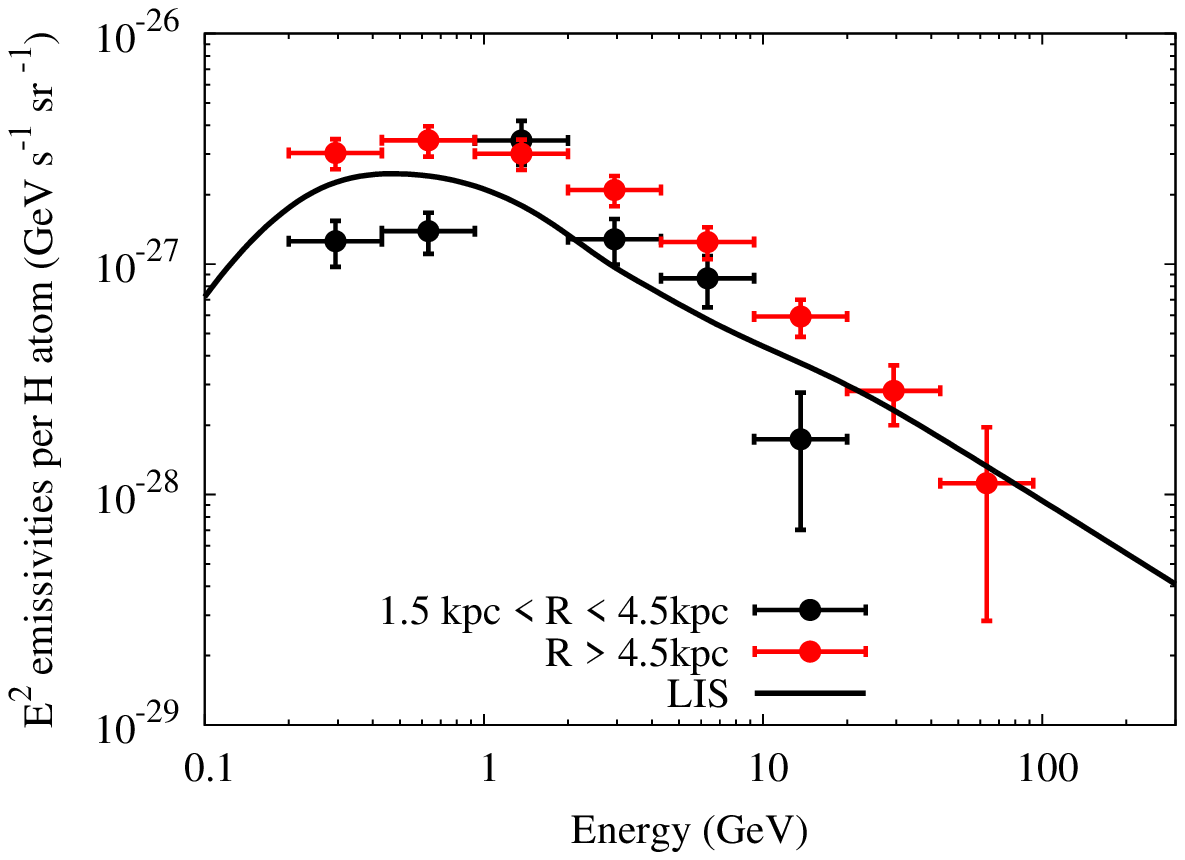}
\includegraphics[width=0.45\hsize]{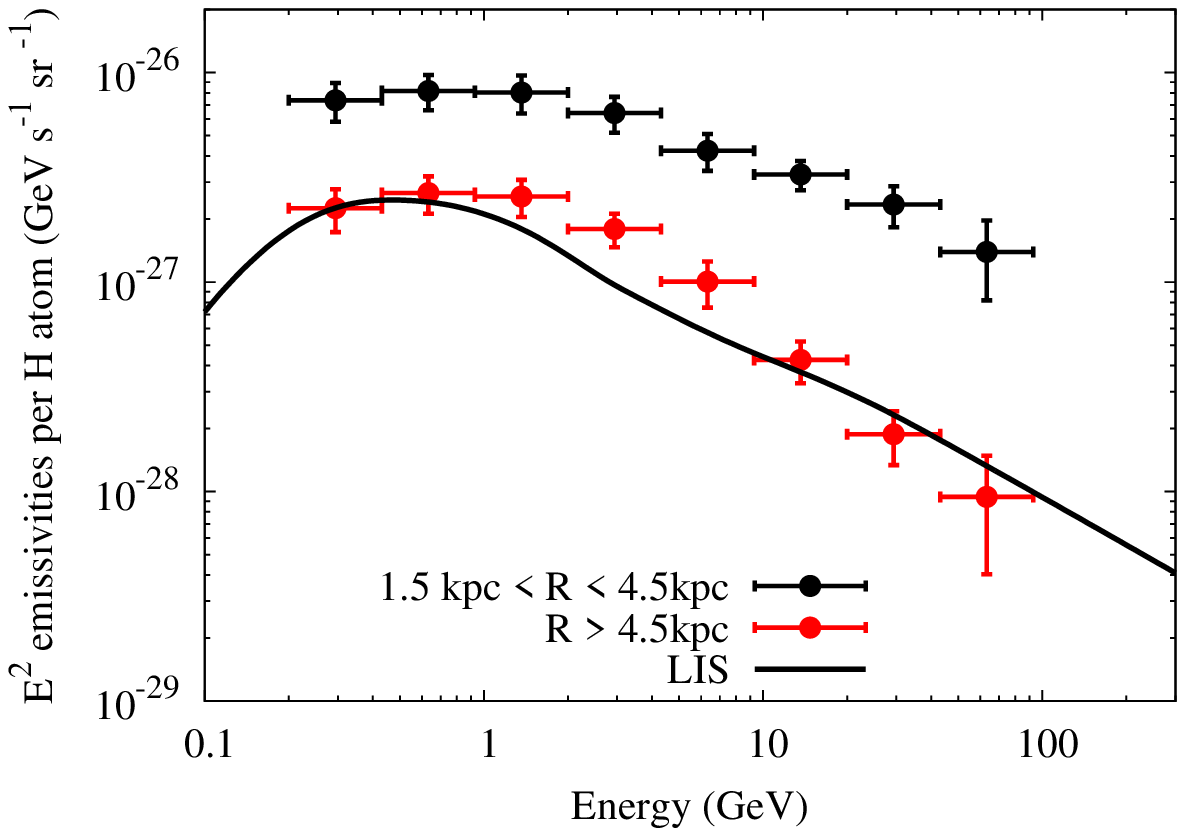}
\caption { \gray emissivities per H atom of diffuse \gray emission associated with the two gas rings in  gas +\ion{H}{ii} models for region I (left panel) and region II (right panel), respectively. }
\label{fig:sedgash2}
\end{figure*}

The TS values of the \ion{H}{ii} templates in the Fermi model are much smaller than those in the other two models. This is not unexpected, since in the standard Fermi diffuse background model the artificial "patches" are added to absorb the residual in the Galactic plane \footnote{https://fermi.gsfc.nasa.gov/ssc/data/access/lat/BackgroundModels.html}, and the standard diffuse model is not suggested to study the diffuse emission regarding the interstellar medium.  And the IC contribution is also fixed. Thus, in the spectral analysis, we used only the gas and dust templates described above. 
We then  divided the energy range 0.2 GeV - 200 GeV into nine logarithmically spaced energy bins and extracted the spectral energy distribution (SED) of diffuse \grays in these regions via the maximum likelihood analysis in each energy bin.
The uncertainties include 68\% statistical errors for the energy flux  and  systematic errors due to the uncertainties in LAT effective areas.
The results for the dust model and dust+\ion{H}{ii} models, as well as gas model and gas+\ion{H}{ii} models,  are shown in  Fig.\ref{fig:spedust}, Fig.\ref{fig:seddust}, Fig.\ref{fig:sedgas}, and  Fig.\ref{fig:sedgash2},  respectively.

In  Fig.\ref{fig:seddust}, Fig.\ref{fig:sedgas}, and Fig.\ref{fig:sedgash2}, we have divided the SEDs with their corresponding gas column densities derived from Sect.\ref{sec:Gas} to obtain the \gray emissivities per H atom, which is proportional to the CR density. The derived emissivities are compared with the one predicted by using the local interstellar spectrum (LIS) of CRs (black line, \citealt{Casandjian15}). We found that in the dust model without including  \ion{H}{ii} templates,  the derived CR spectrum from the dust model of each region is significantly harder than the LIS, which is consistent with the results for the inner Galaxy in \citet{yang16} and \cite{fermi_diffuse}. 
However, for the dust+\ion{H}{ii} models, the derived \gray emissivities associated with the dust templates are in good agreement with the local emissivities, which imply similar CR spectra in these regions and  the solar neighborhood. Meanwhile, the derived emissivities associated with the \ion{H}{ii} component are much harder, with an index of about $-2.1$. 

For the gas model without the \ion{H}{ii} templates, the derived \gray emissivities for the inner rings  (below 4.5 kpc) are  significantly higher than the LIS value and the outer rings (above 4.5 kpc), while the spectra in the inner rings is significantly harder. These results are also in good agreement with the results of \citet{yang16} and \cite{fermi_diffuse}. When including \ion{H}{ii} templates, for region I the \gray emissivities for both inner and outer gas rings are consistent with the LIS value in both the spectra and normalizations. For region II, the \gray emissivities' spectral indices  in both rings are consistent with the LIS value, while the normalization in the inner ring is higher. 
The \gray emissivities from \ion{H}{ii} templates are also significantly harder in both regions and are consistent with the results derived from dust+\ion{H}{ii}models in both spectra and normalizations.

It should be noted that the normalization of the derived emissivities associated with \ion{H}{ii} component is uncertain. This is because the ionized gas only contributes a small portion to the total gas column in our Galaxy. Moreover, it is possible that the CRs associated with ionized gas also illuminate other gas components. In  Fig.\ref{fig:seddust} and Fig.\ref{fig:sedgash2}, we assumed only \ion{H}{ii} gas are responsible for the \grays associated with the \ion{H}{ii} templates.  Thus, the emissivities derived here can be regarded as the upper limits.

\section{Discussion}
\label{sec:discussion}

In this work, we found a hard \gray component associated \ion{H}{ii} in two regions in the Galactic plane. The derived intensity and spectrum of such a \gray component are stable when adopting different gas tracers and IC models. Including such a component will also alter the \gray emissivities associated with the gas in the Galactic plane. This can be seen in both the dust and gas models in our analysis. In the dust model, the inclusion of the \ion{H}{ii} component will deduce a similar \gray emissivities in the whole line of sight as the local value, which implies that the average CR density and spectra in the line of sight are similar to the local CR density and spectrum. Such a conclusion is different from those in \citet{fermi_diffuse} and \citet{yang16}. Moreover, in the gas model, the gases are split into two rings with galactocentric radii below and above 4.5 kpc. Without the \ion{H}{ii} component, we reproduced the results in \citet{fermi_diffuse} and \citet{yang16};  thus, the \gray emissivities per H atom are higher in the inner ring and reveal a significantly harder spectrum. Including the \ion{H}{ii} template, we found the \gray emissivities in both rings become consistent with the local value  for region I.  For region II, the spectra are similar to the local value, while for the inner rings the normalizations are still significantly higher. However, we note that the normalization of \gray emissivities per H atom is inversely proportional to the $X_{co}$ conversion factor, which is fixed in our analysis. Indeed, $X_{co}$ depends on the metallicity and can be different in different parts of the Galaxy \citep{fermi_diffuse}.

\begin{figure*}[ht]
\centering
\includegraphics[width=0.45\hsize]{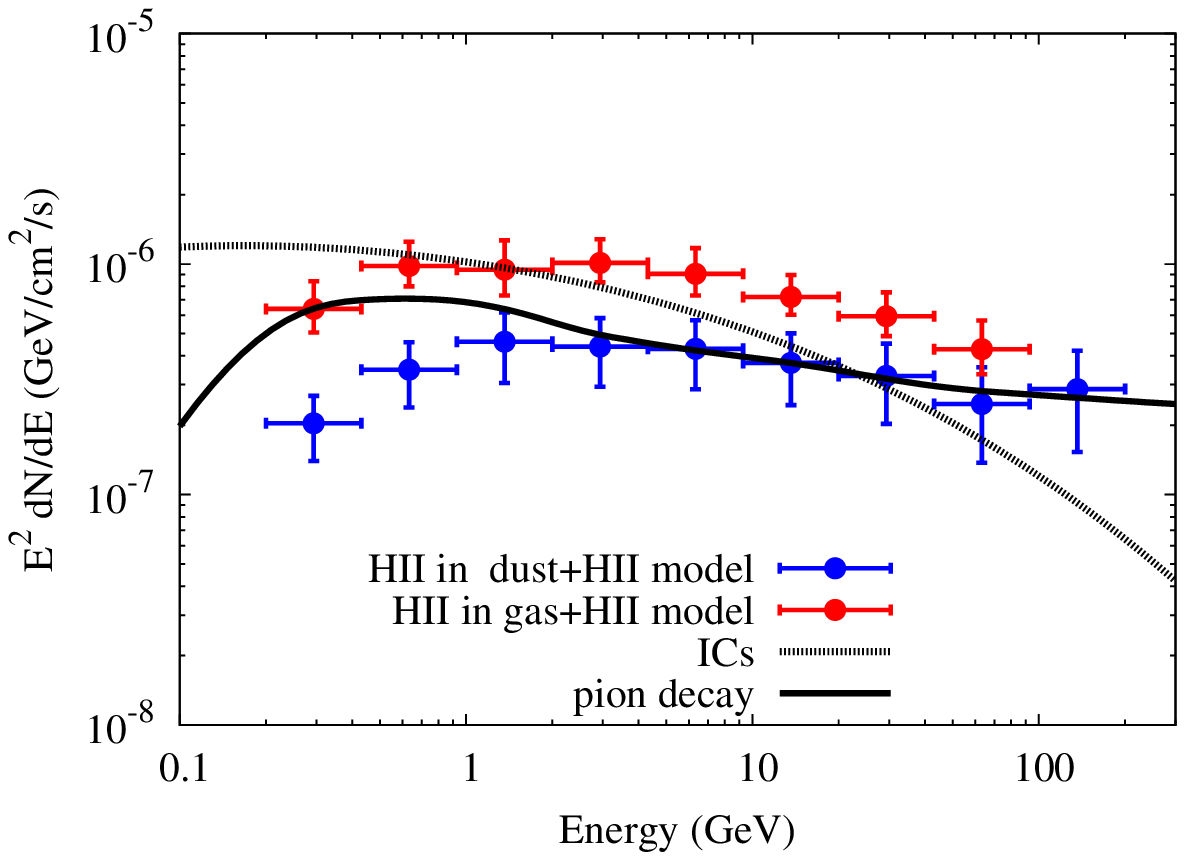}
\includegraphics[width=0.45\hsize]{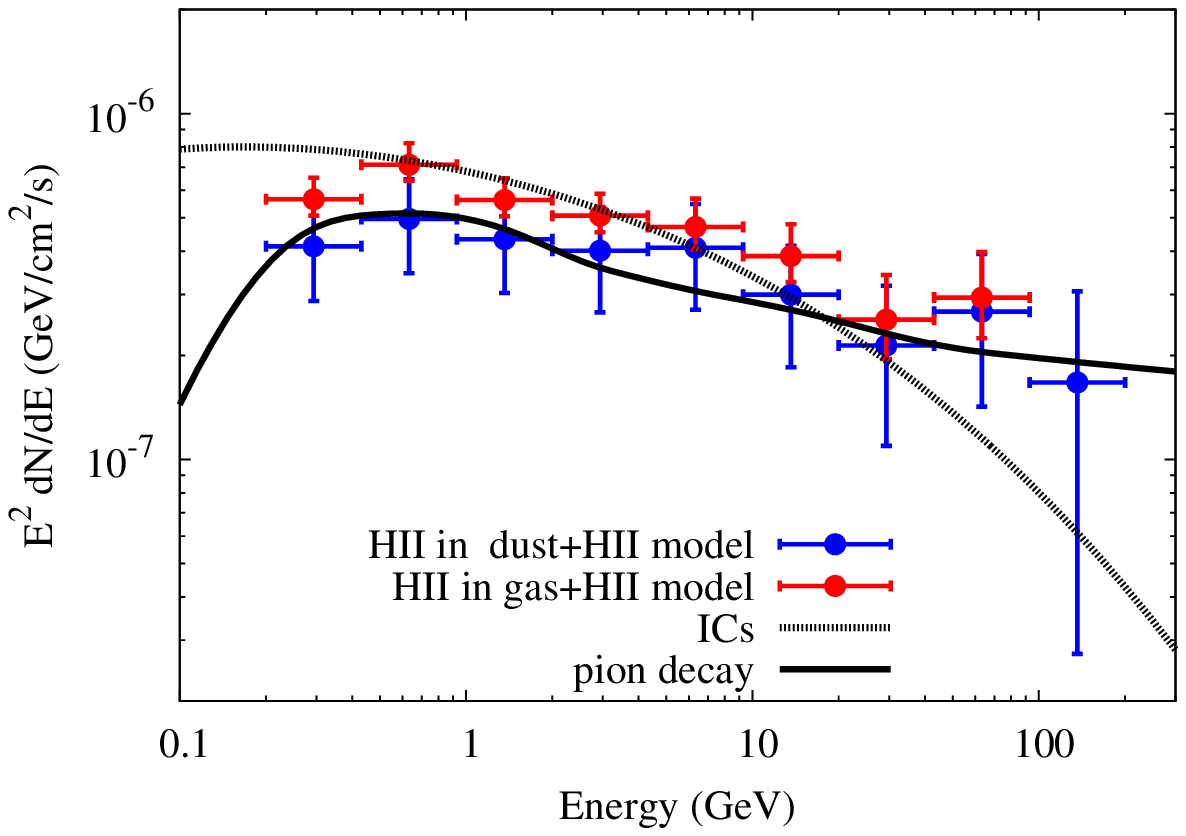}
\caption { \gray flux  associated with \ion{H}{ii} gas  obtained for dust+\ion{H}{ii}  models (blue) and gas+\ion{H}{ii}  models (red) for region I (left panel) and region II (right panel), respectively. Also plotted are the possible contributions from IC and pion-decay processes.  The details can be found in Sect.\ref{sec:discussion} .}
\label{fig:sedh2}
\end{figure*}

To conclude, we found a robust hard \gray emission component associated with the \ion{H}{ii} gas, and the \gray emissions associated with other gas reveal a significantly softer spectrum with a similar spectral shape. In the following discussion, we label the \gray emission and corresponding CR population related to the dust and gas templates and \ion{H}{ii} templates as "soft" and "hard" components, respectively. 
As for the hard component, the absolute normalization can not be determined decisively, as we discussed above.
By using the upper limit of the emissivities in the most extreme case, the hard component is only produced in the \ion{H}{ii} region, the derived CR density can be 100 times higher than the LIS value (see Fig.\ref{fig:seddust} and Fig.\ref{fig:sedgash2}). Such enhancement of CR densities is expected near CR sources such as SNRs and YMCs. The dimension of the regions with enhanced CR density can be of more than 100~pc \citep{aharonian19}. Thus in this case, the hard \gray component associated with \ion{H}{ii} gas can be interpreted as the enhanced \gray emission in the vicinity of sources.   To illustrate such a scenario, we plotted in Figs.\ref{fig:sedh2} the \gray spectrum assuming a power-law parent proton spectrum with an index of $-2.2$, which is similar to the CR spectrum derived in the vicinity of YMCs \citep{aharonian19}.  On the other hand, the hard \gray emission can also have a leptonic origin, especially considering the enhanced radiation fields in these \ion{H}{ii} regions. The interstellar radiation fields (ISRFs), which provide the low-energy photon targets for the IC process, can be divided into three components in our Galaxy \citep{Popescu17}, including the optical and UV fields from the starlight, the infrared fields from the dust emission, and the cosmic microwave backgrounds. In the \ion{H}{ii} regions, the ionizing massive stars will increase the optical and UV fields significantly and thus produce additional IC emissions. We calculated such a component assuming the electron spectrum is the same as the electron spectrum measured in the solar neighborhood and from radio observations \citep{strong11}, and the optical and UV components are described by a grey-body spectrum with a temperature of $5000 ~\rm K$. The derived IC \gray fluxes are also shown in Fig.\ref{fig:sedh2}. In the calculated  \gray flux from both pion decay and the IC component, the total normalization is rather arbitrary, since both the average gas density and the energy densities of the enhanced optical and UV radiation fields are unknown. But the derived spectral shapes already reveal significant differences. The IC components, due to the Klein-Ninisha effects, are softened above dozens of GeVs. The current data cannot rule out the leptonic origin, but further observations above 100 GeV may distinguish two such possibilities. If there is another hard electron spectrum in these \ion{H}{ii} regions, the IC component can be harder than the curve calculated here, but the Klein-Nishina effects for the optical and UV components will inevitably play a role above several hundred GeV in the \gray spectrum \citep{Popescu17}. Thus, the observations above several hundred GeV may distinguish such two scenarios.

Interestingly, a two-component CR model was used in \citet{yang19} to explain the hardening of CR spectrum above 200 GV, as well as the increasing positron/electron and antiproton/proton ratios.  In such a scenario, CRs are injected from two populations of sources with different spectral properties (more specifically, a 1st source population with soft spectra and a 2nd source population with hard spectra). For the second source population, CRs accumulate the grammage of about $1~\rm g/cm^2$ in the vicinity of the sources. Accordingly, they should be effective \gray emitters. \citet{yang19} calculated diffuse \gray emissions associated with the second (hard)  source population and found it can dominate over the first (soft) source population above about 100~GeV, which is in agreement with our measurements shown in  Fig.\ref{fig:spedust}.  
As mentioned in \citet{yang19}, one possible candidate for the second source population can be YMC. Indeed, the OB star distributions do have a strong peak at about 4~kpc away from the GC \citep{bronfman00}. Therefore, we expect a strong enhancement of the hard \gray emissions associated with these sources in this region, which can naturally explain the high energy density and hard spectrum of CRs derived from the 4-6 kpc rings in our Galaxy \citep{fermi_diffuse, yang16}. 

On the other hand, the similar density and spectral shape of the soft CR component with the LIS reveal the possible CR sea all over the Galaxy. Such a possibility is also unveiled in \citet{aharonian20}, in which the \gray observations on individual molecular clouds are used to measure the CR density in different positions of our Galaxy. Such a uniform CR distribution is not trivial, since if the CR propagation is dominated by diffusion, the spatial distribution of the sources should always leave a trace in the spatial distribution of CRs \citep{strong98}. However, all the promising candidates of CR accelerators, such as SNRs \citep{green15}, pulsars \citep{yusifov04}, and OB stars \citep{bronfman00}, reveal a significant inhomogeneous distribution in our Galaxy. Thus, a larger CR halo or strong re-acceleration may be needed to explain the uniform distribution of the  CRs, which will lead to a significant modification to the current paradigm of CR propagation in our Galaxy.

Finally, we note that  although the fitting of \gray  data is significantly improved by introducing the \ion{H}{ii} gas component, there are still strong residuals (see   Fig.\ref{fig:ts_reg1} and Fig.\ref{fig:ts_reg2}). This is not unexpected if we assume the hard \gray component is from the vicinity of CR sources. Since the CR distribution should be significantly inhomogeneous near the source, we expect no perfect spatial correlation between the gas and the \gray emissions. In this regard, a detailed analysis in each compact \ion{H}{ii} region may be helpful. These regions are often extremely crowded and different sources can be confused and overlap in the line of sight \citep{liub2019}. However, as shown in \citet{aharonian19} these sources can also be PeVatrons. If this is the case, the corresponding \gray emission should be higher than 10~TeV. In this energy band, the background due to the diffuse \gray emission associated with soft CR components and individual \gray sources such as pulsars, SNRs, and pulsar wind nebulae can be strongly suppressed. The extensive shower arrays such as LHAASO \citep{lhaaso2016} and HAWC \citep{hawc}, as well as the Cerenkov telescope arrays  such as H.E.S.S., MAGIC, VERITAS, and CTA would be ideal instruments to explore these regions.

\section*{Acknowledgements}
Bing Liu is supported by  the NSFC under grant 12103049.
Ruizhi Yang is supported by  the NSFC under grants 11421303, 12041305 and  the national youth thousand talents program in China. 
\bibliographystyle{aa}

\begin{appendix}

\section{Residual significance (signal-to-noise, S/N) maps for regions I and II }

\begin{figure*}[ht]
\centering
\includegraphics[width=0.33\textwidth]{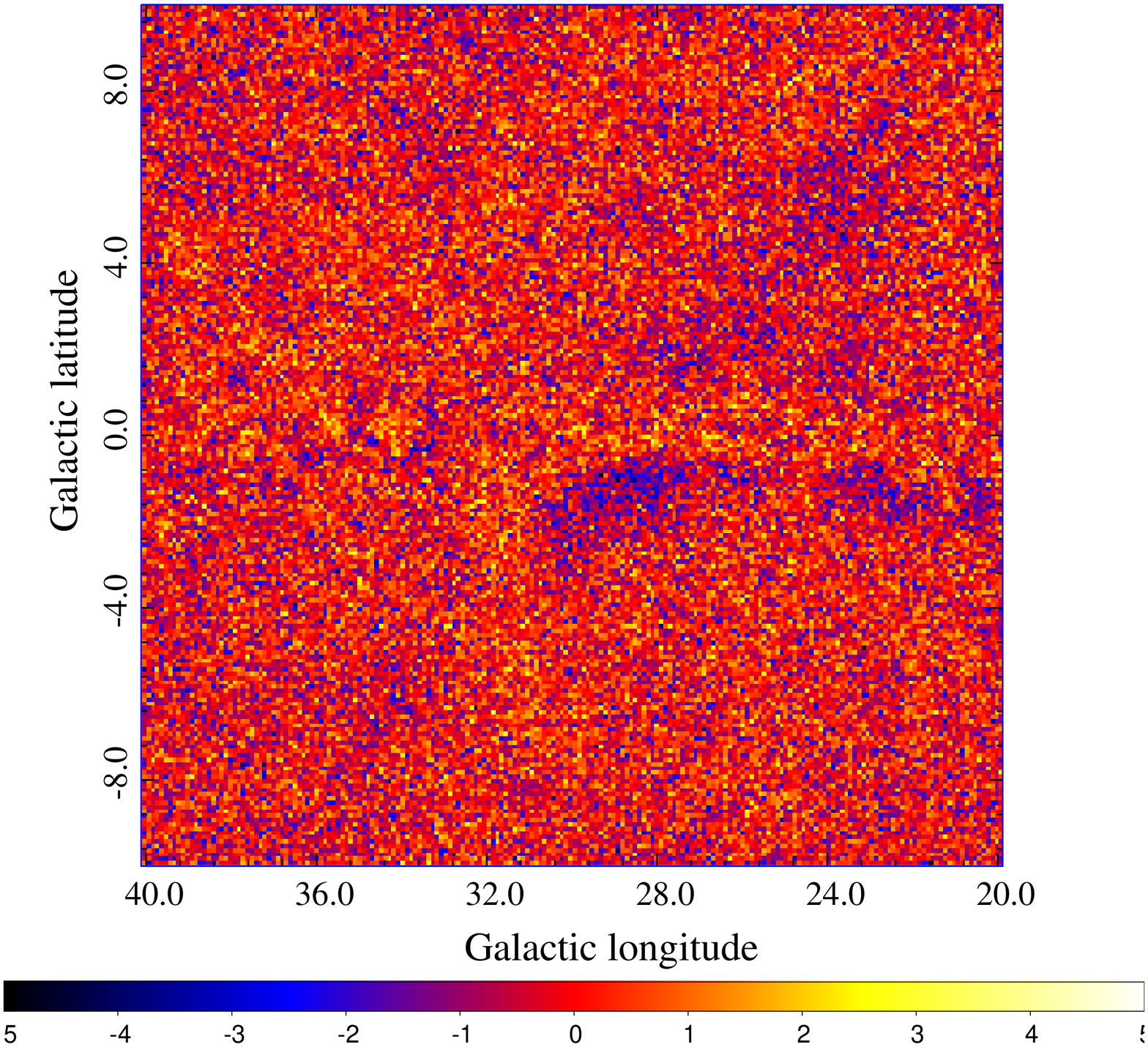}
\includegraphics[width=0.33\textwidth]{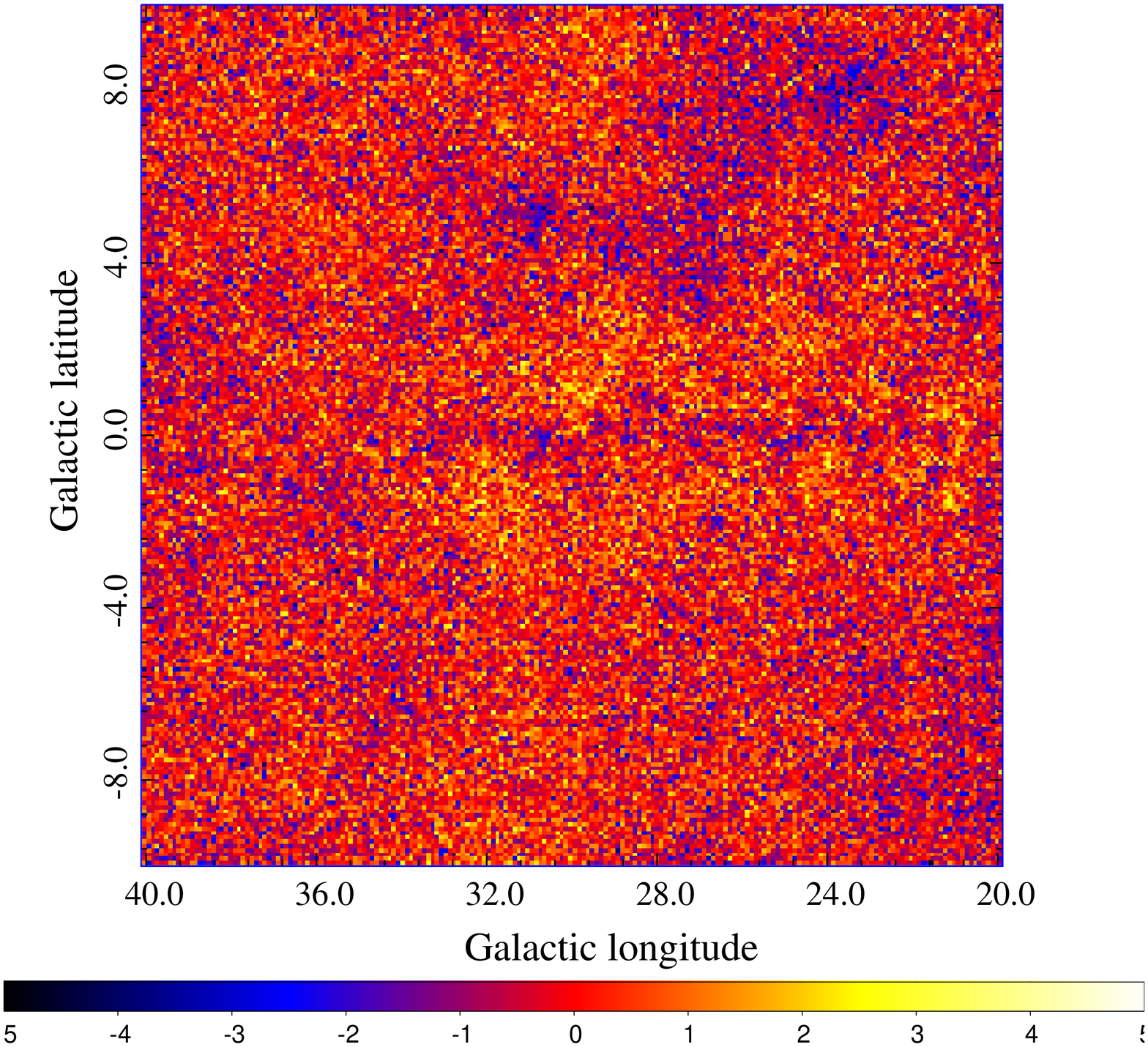}
\includegraphics[width=0.33\textwidth]{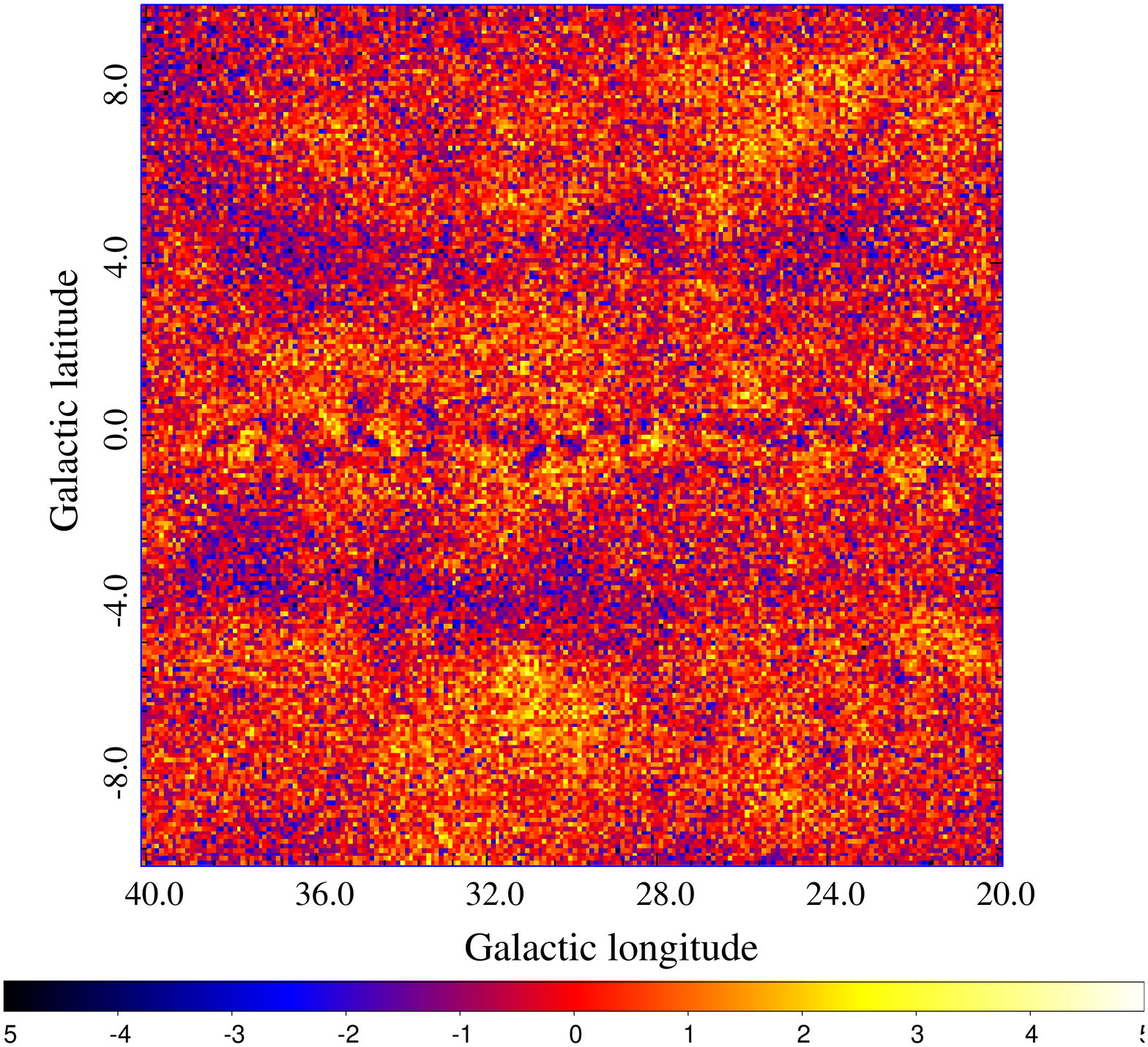}\\
\includegraphics[width=0.33\textwidth]{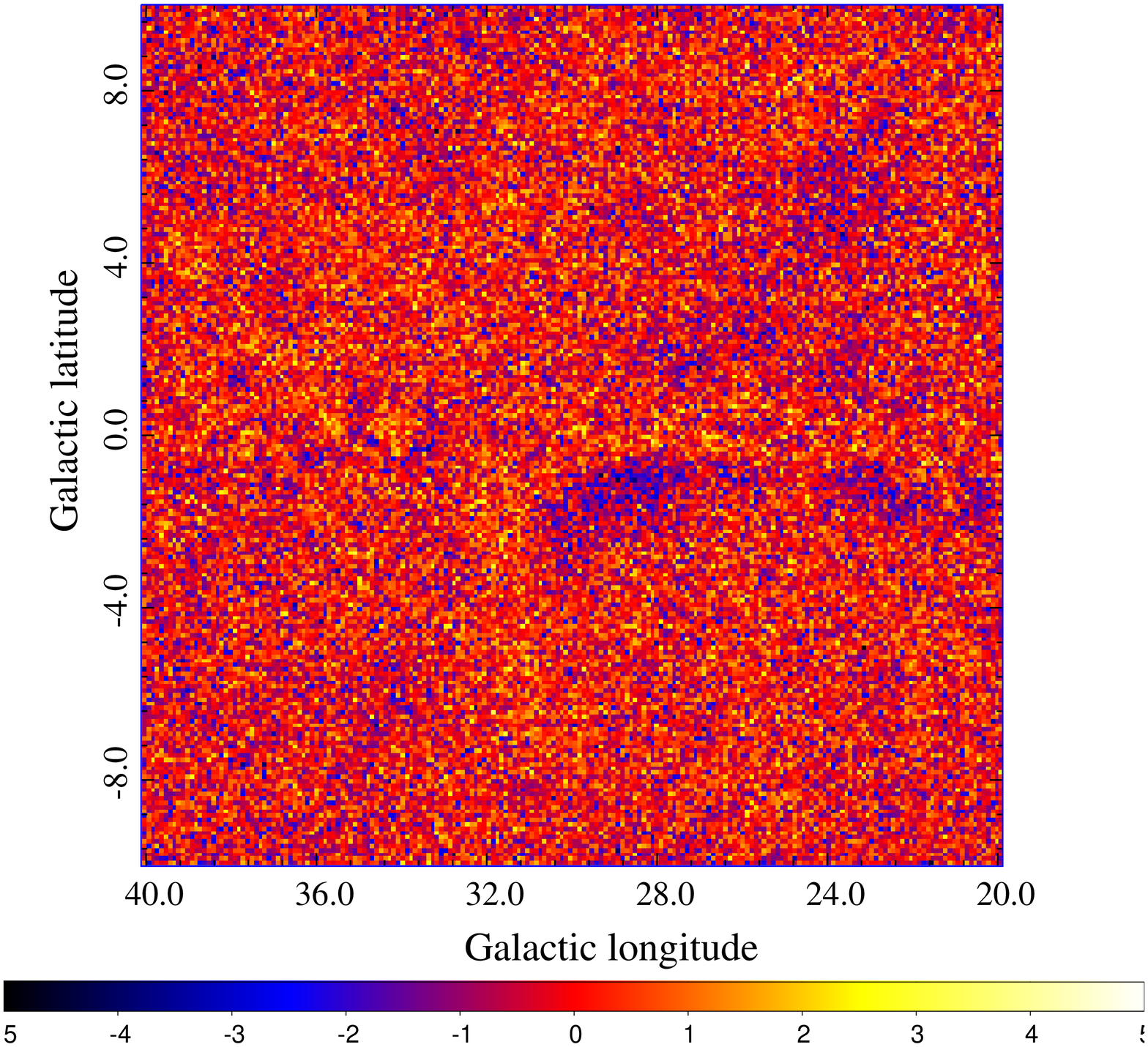}
\includegraphics[width=0.33\textwidth]{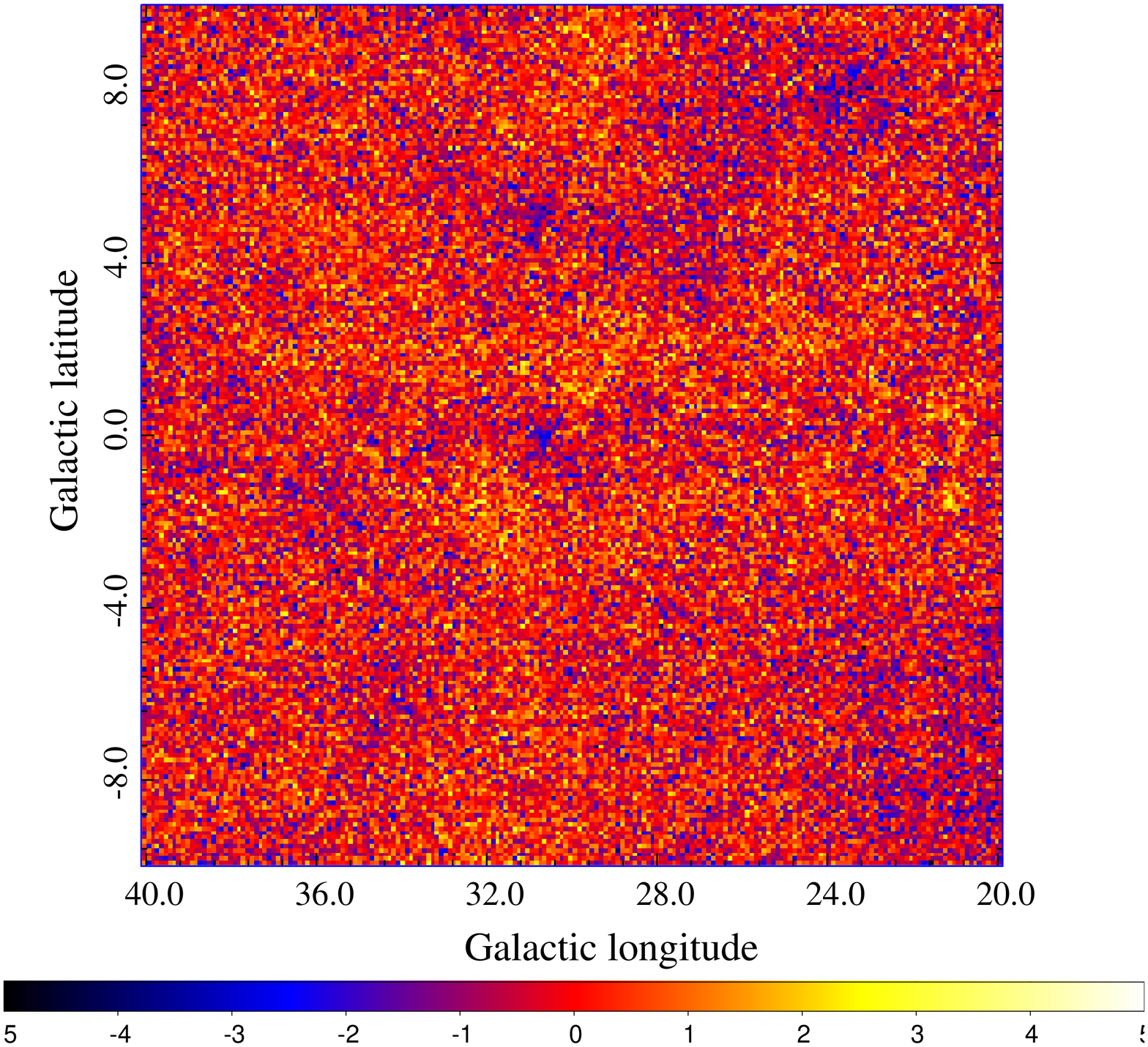}
\includegraphics[width=0.33\textwidth]{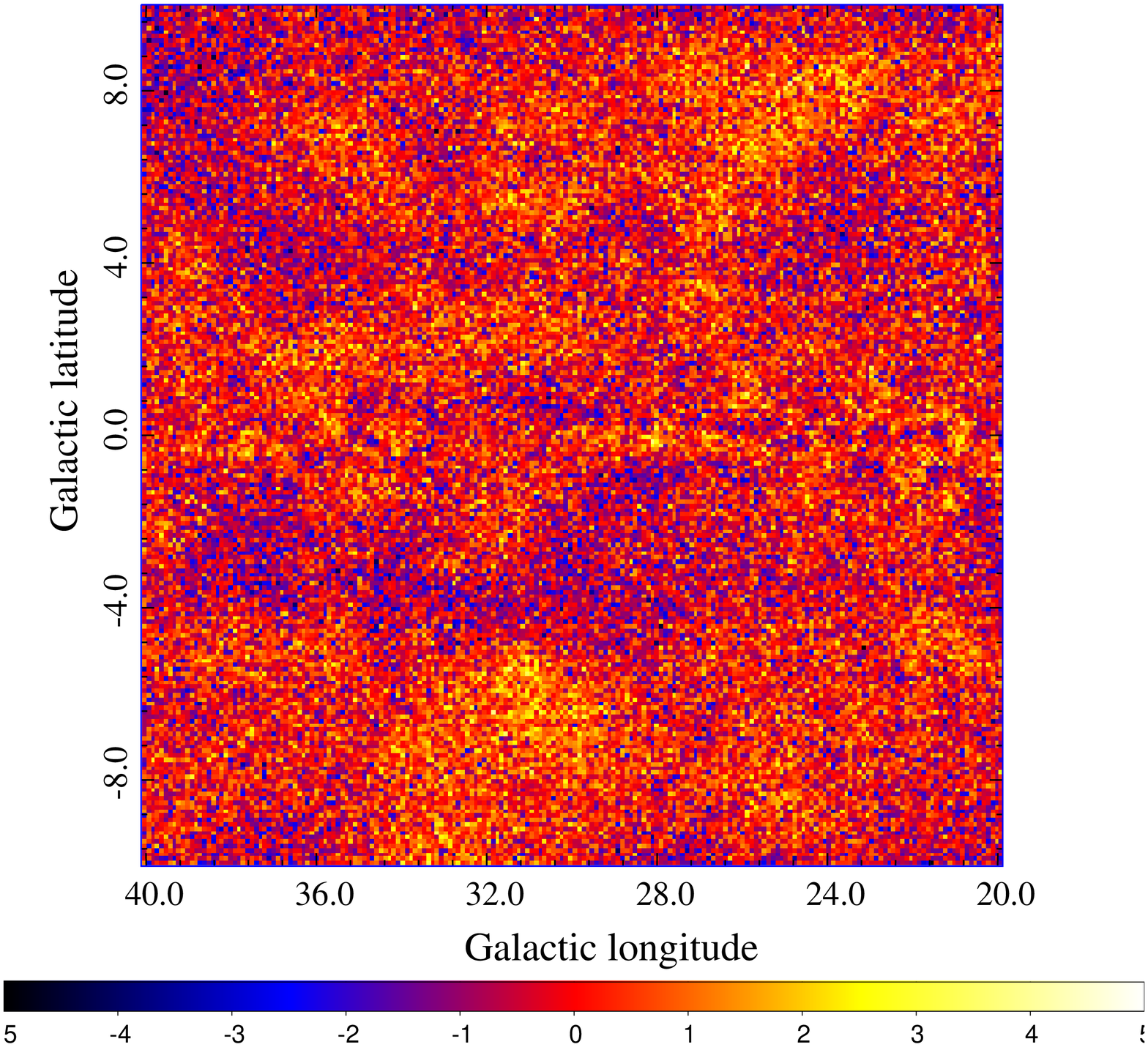}\\
\includegraphics[width=0.33\textwidth]{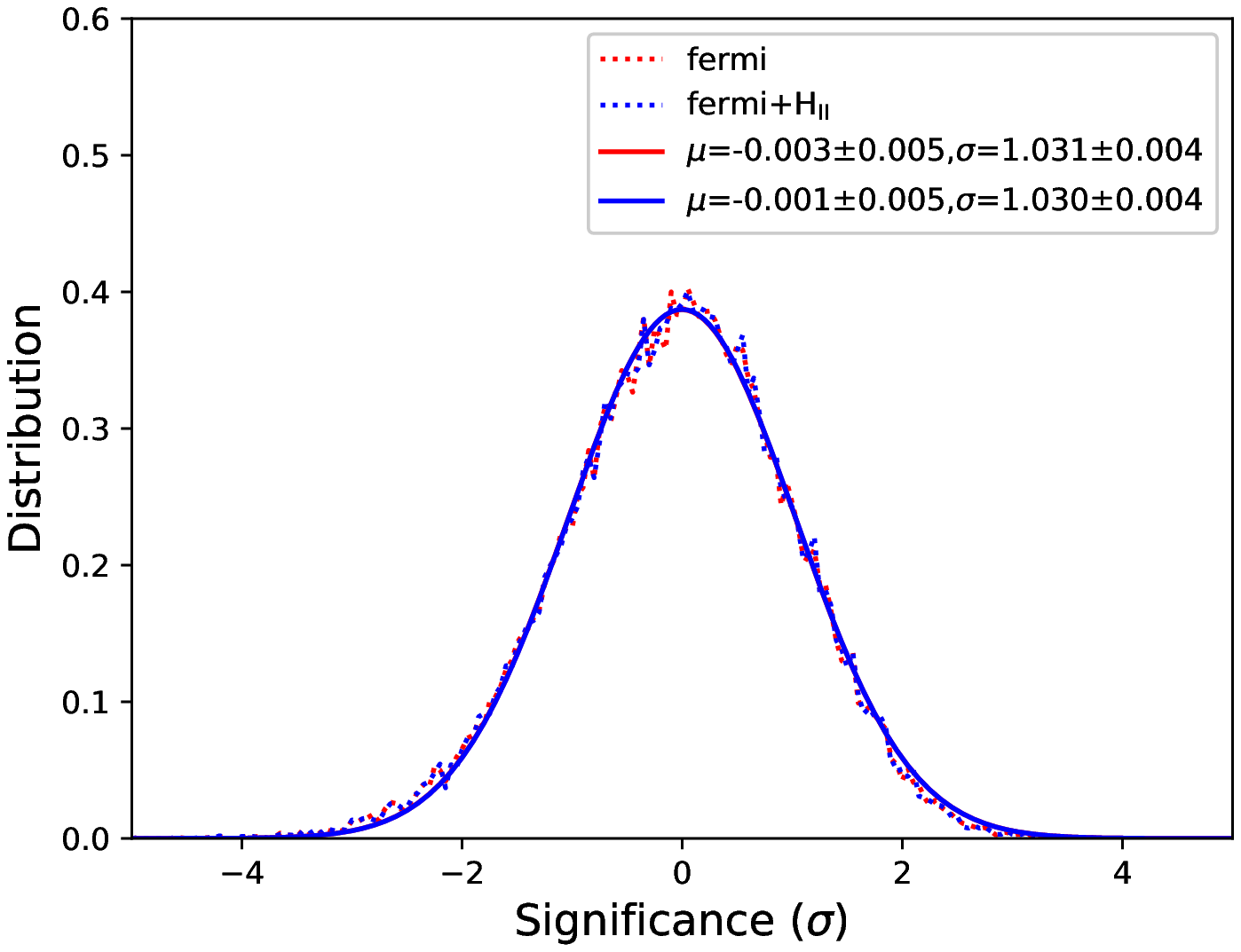}
\includegraphics[width=0.33\textwidth]{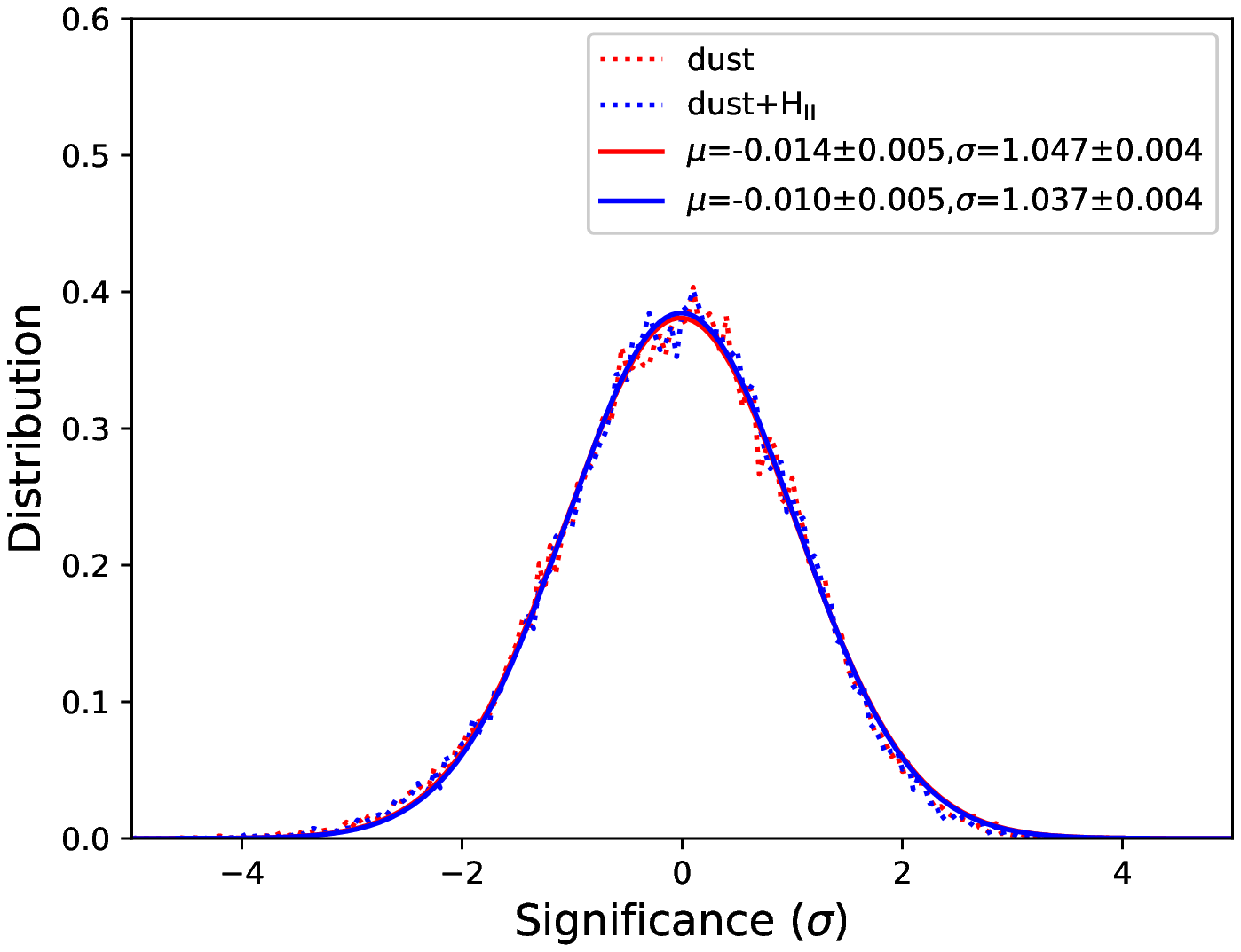}
\includegraphics[width=0.33\textwidth]{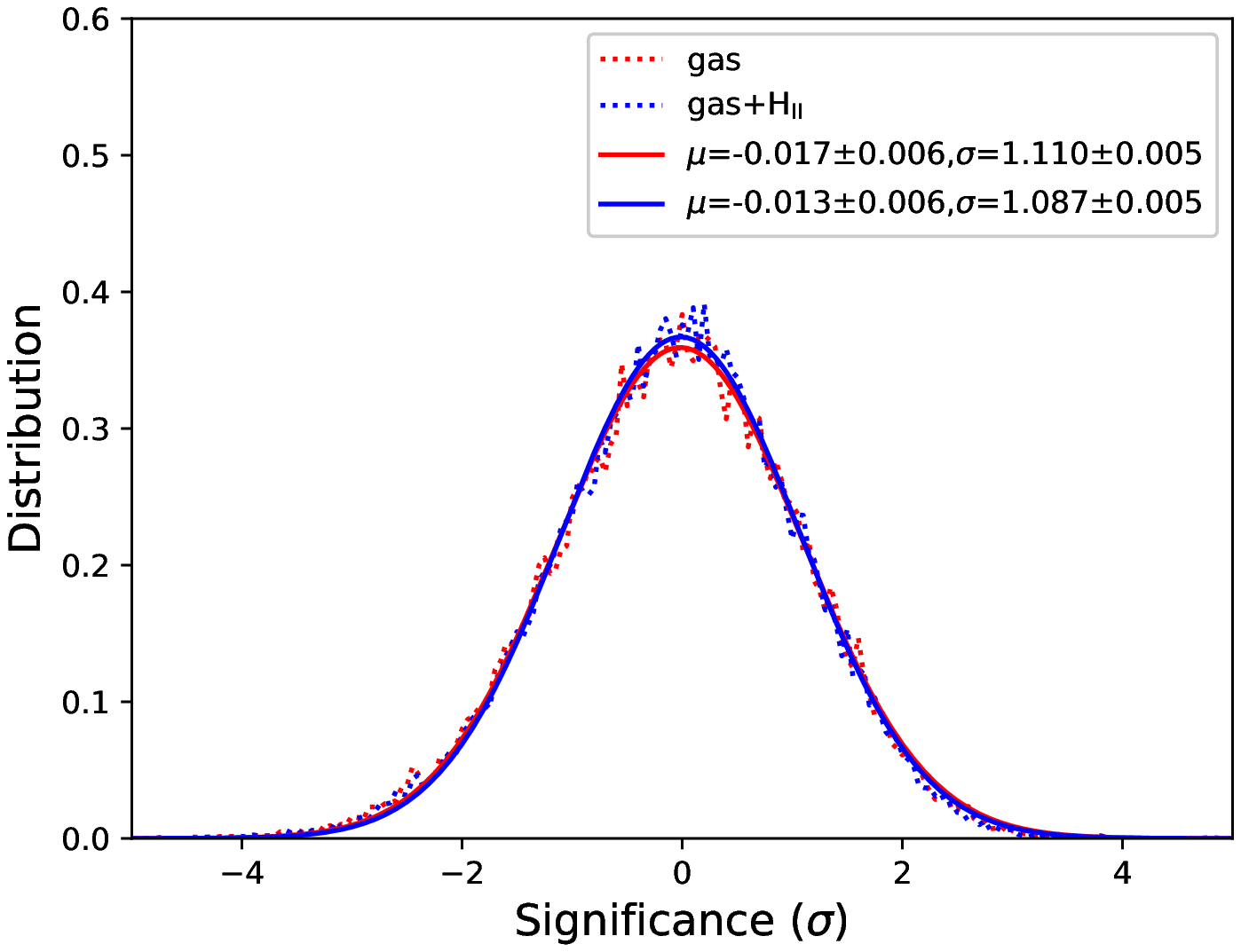}\\
\caption { Comparison of the residual significance (signal-to-noise, S/N) maps in the energy range of 0.2--200 GeV for models without an \ion{H}{ii} component (top) and with an \ion{H}{ii} component (middle) of region I. The The left panels show the results of Fermi and Fermi$+$\ion{H}{ii} models, the middle panels show the results of  dust and dust$+$\ion{H}{ii} models, and the right panels illustrate the results of gas and gas$+$\ion{H}{ii} models. The color scales are from $-5 \sigma$ to $5 \sigma$. The bottom panel  show the distribution of significance (bin size is 0.05 $\sigma$ of each model without or without the HII component, in which each solid line illustrates the Gaussian function fitted to the corresponding distribution and $\mu$ and $\sigma$ represent the mean and the standard deviation of each fit. 
} 
\label{fig:ts_reg1}
\end{figure*}

\begin{figure*}[ht]
\centering
\includegraphics[width=0.33\textwidth]{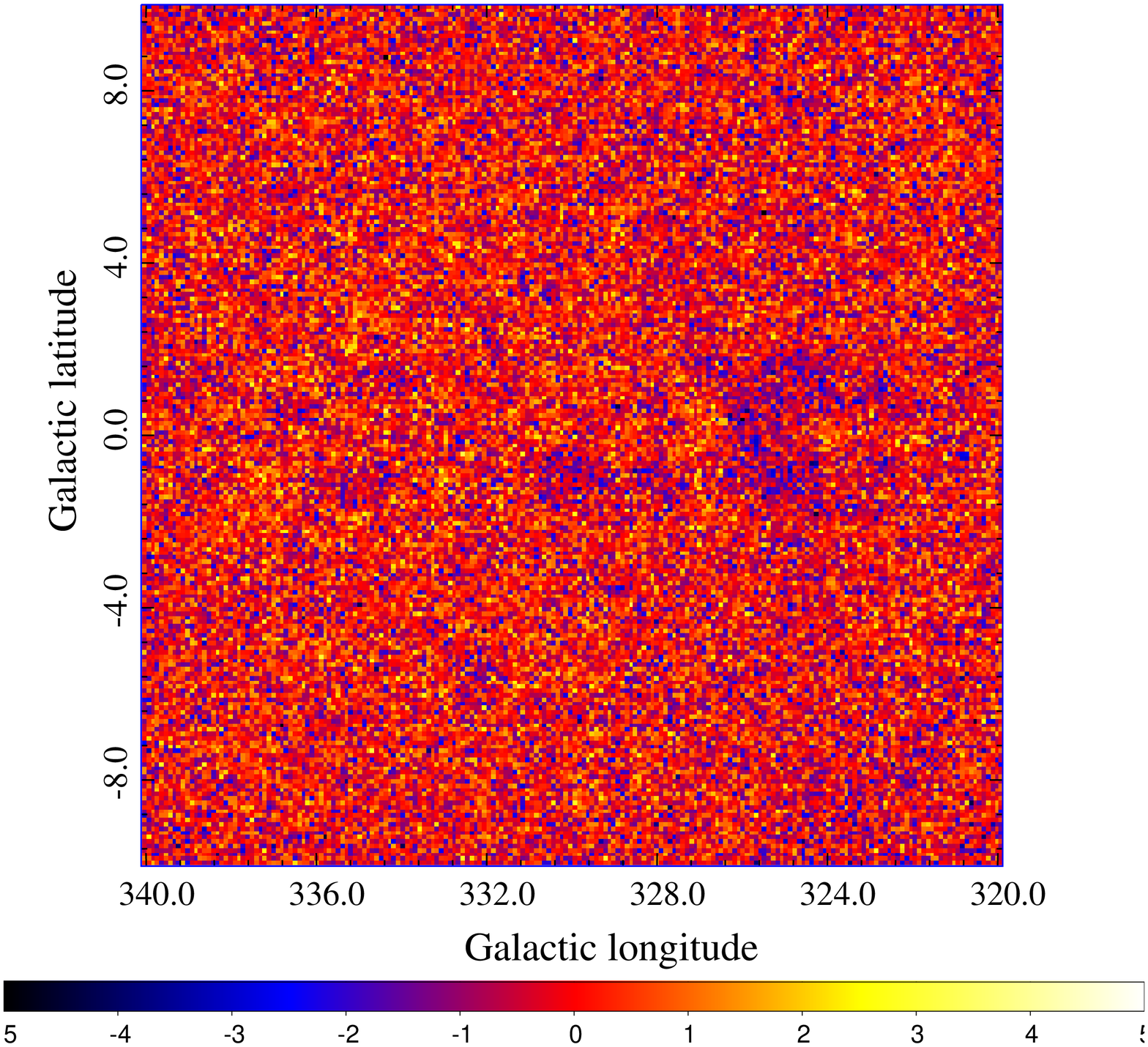}
\includegraphics[width=0.33\textwidth]{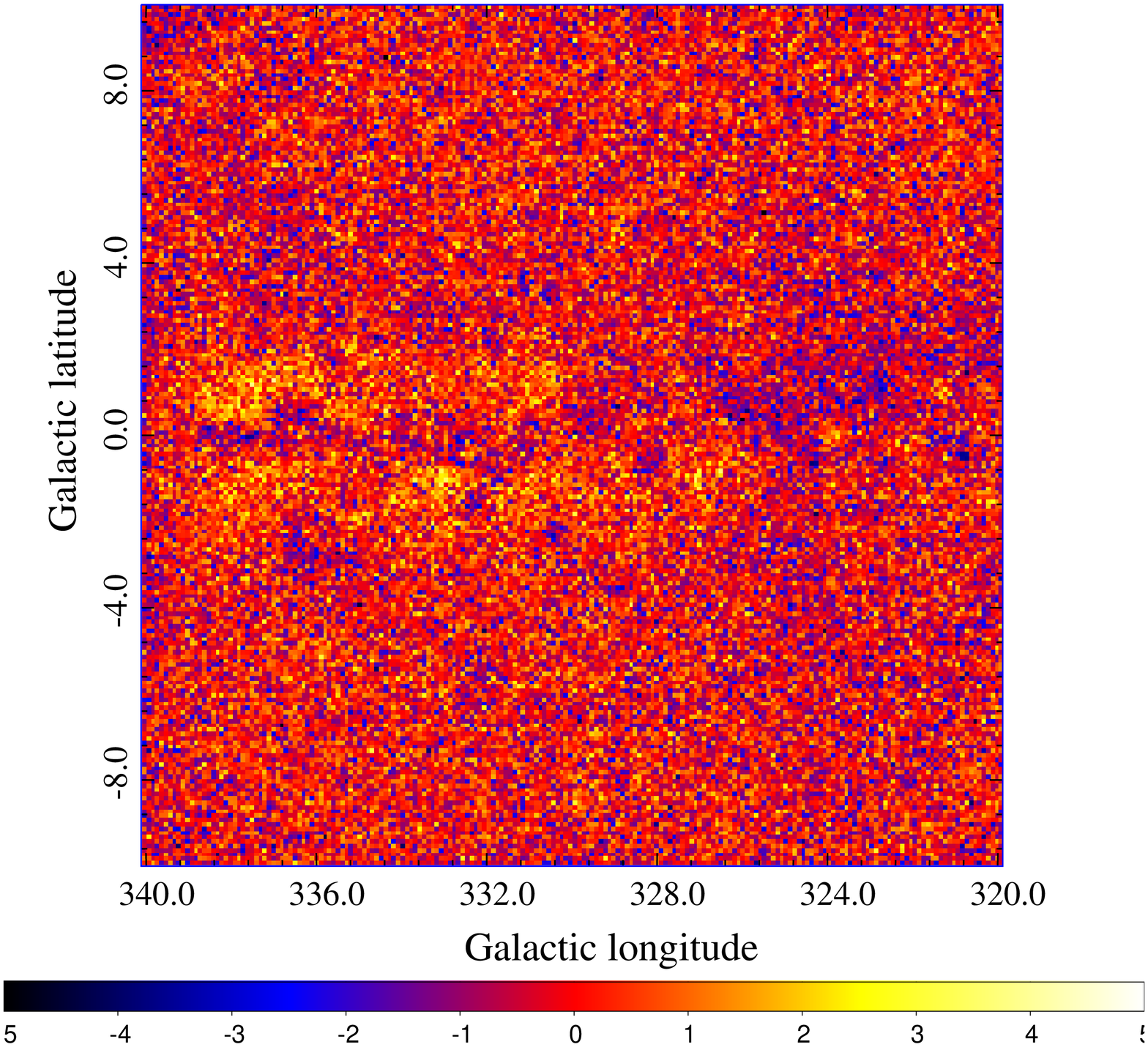}
\includegraphics[width=0.33\textwidth]{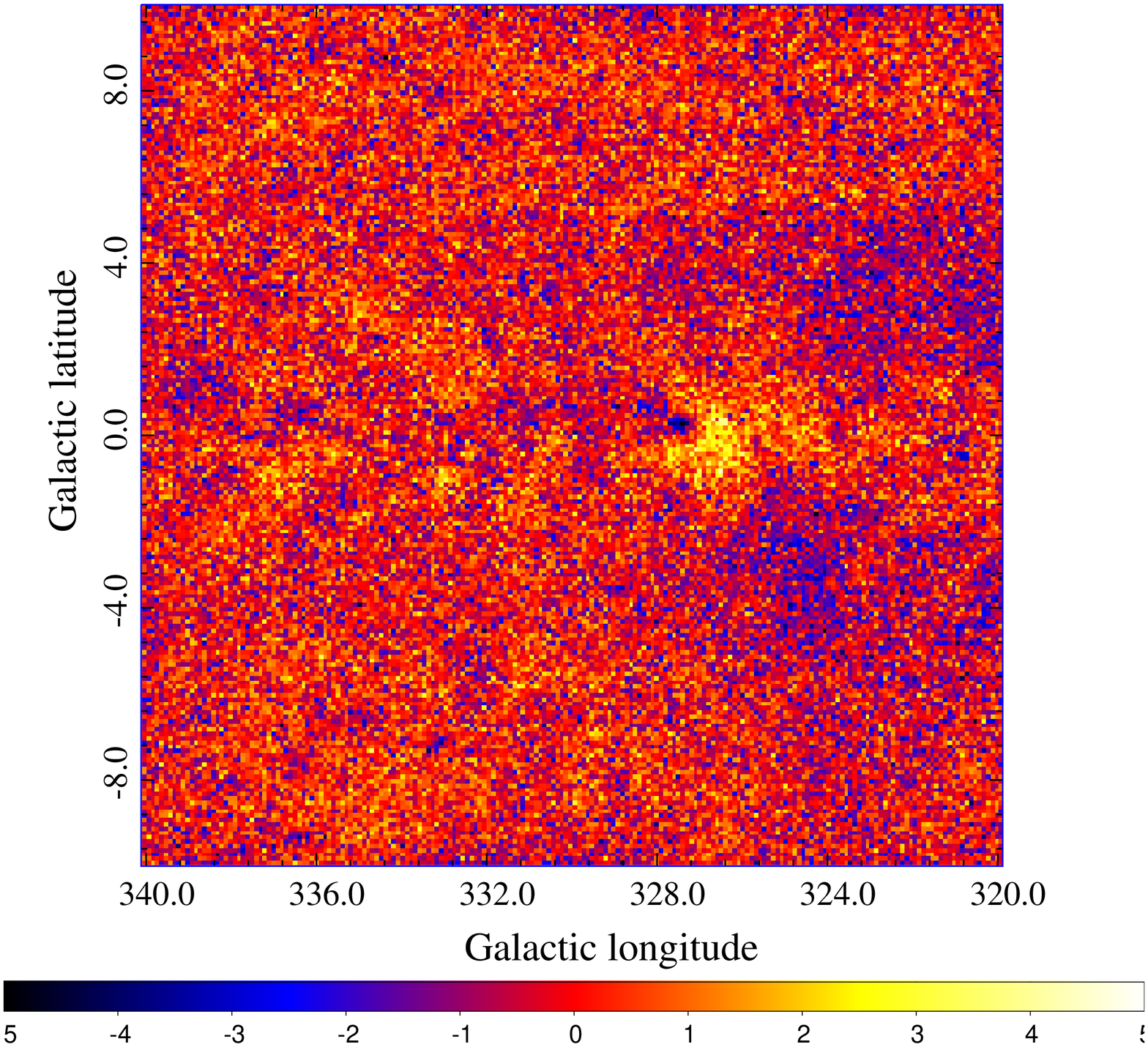}\\
\includegraphics[width=0.33\textwidth]{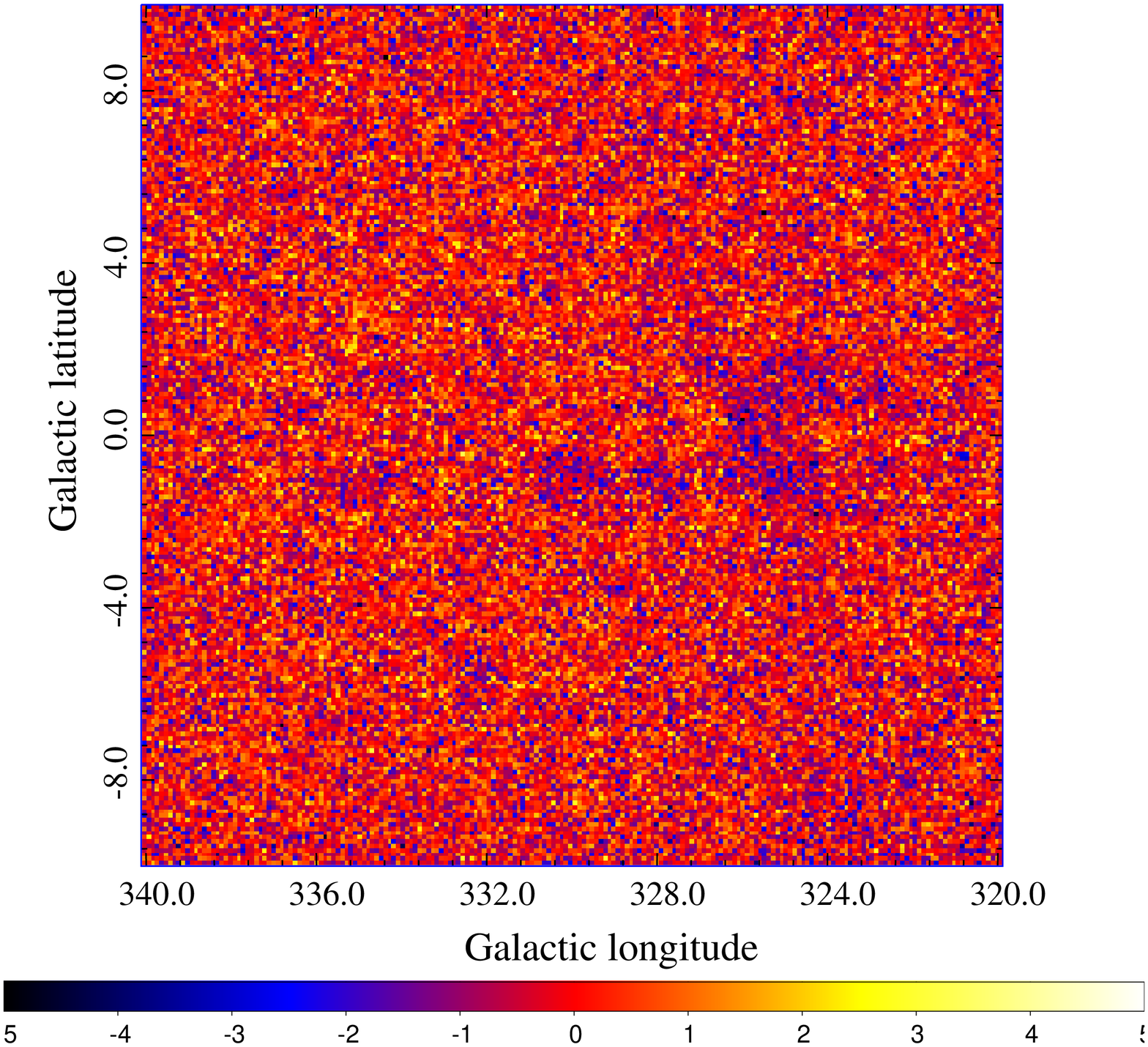}
\includegraphics[width=0.33\textwidth]{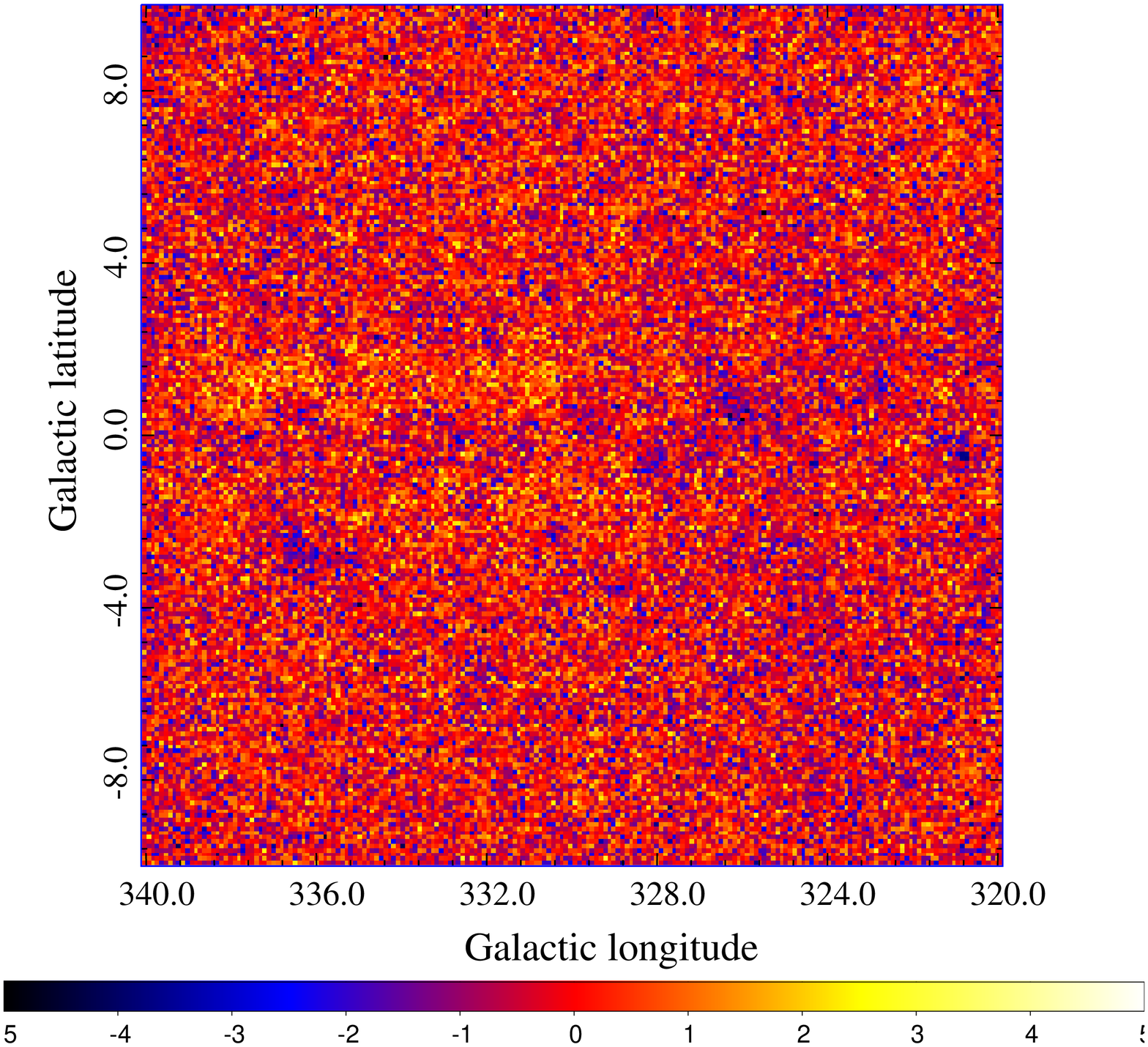}
\includegraphics[width=0.33\textwidth]{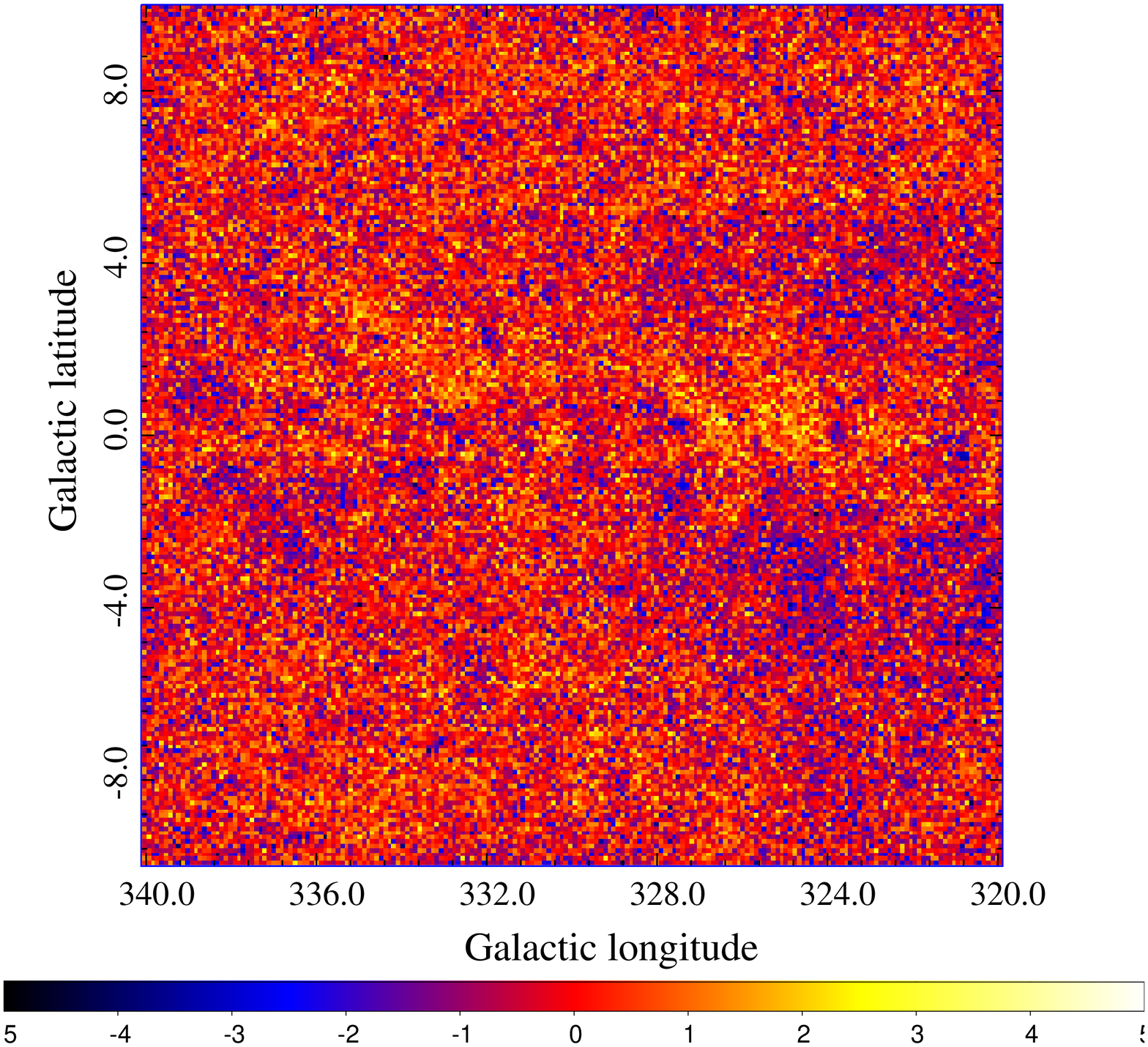}\\
\includegraphics[width=0.33\textwidth]{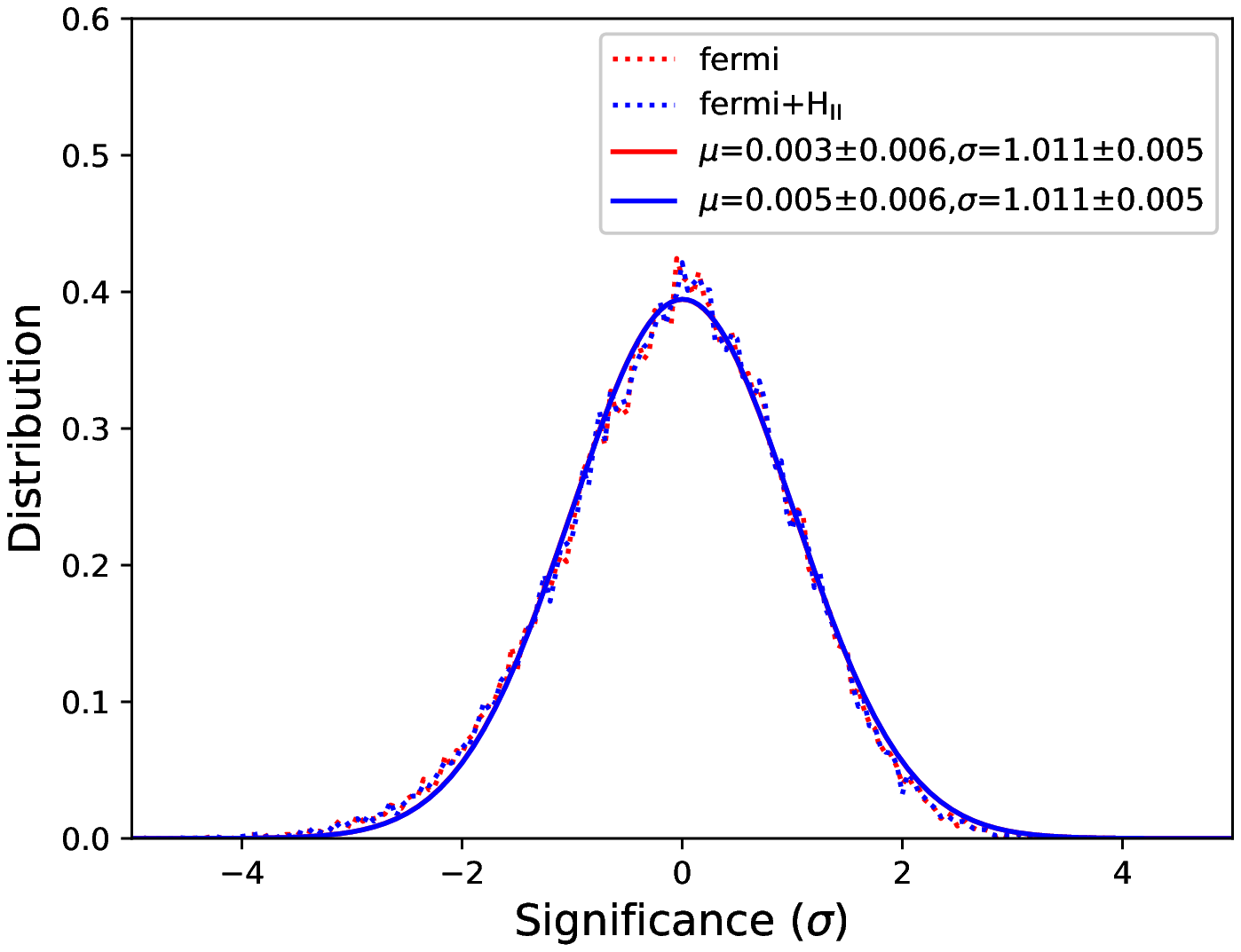}
\includegraphics[width=0.33\textwidth]{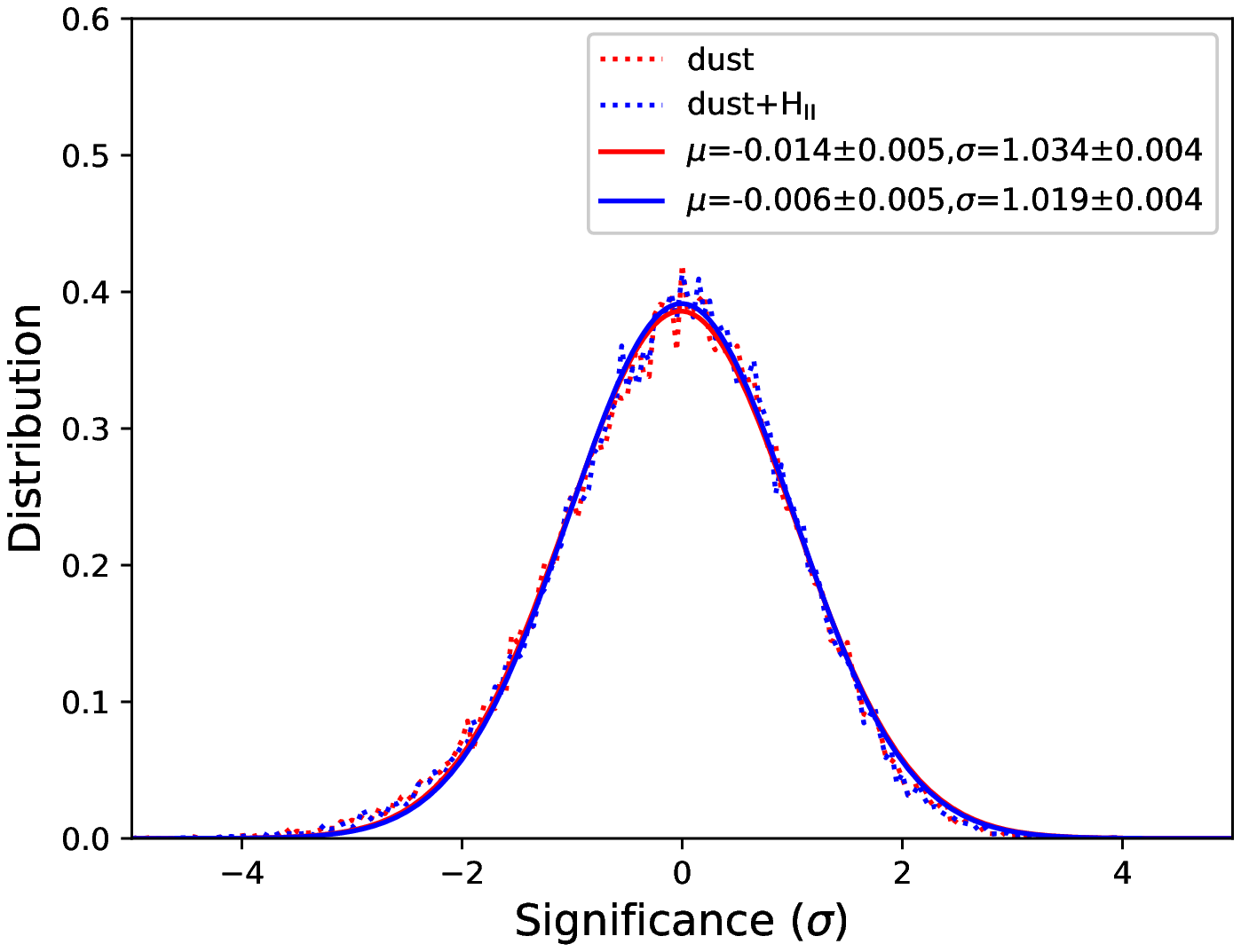}
\includegraphics[width=0.33\textwidth]{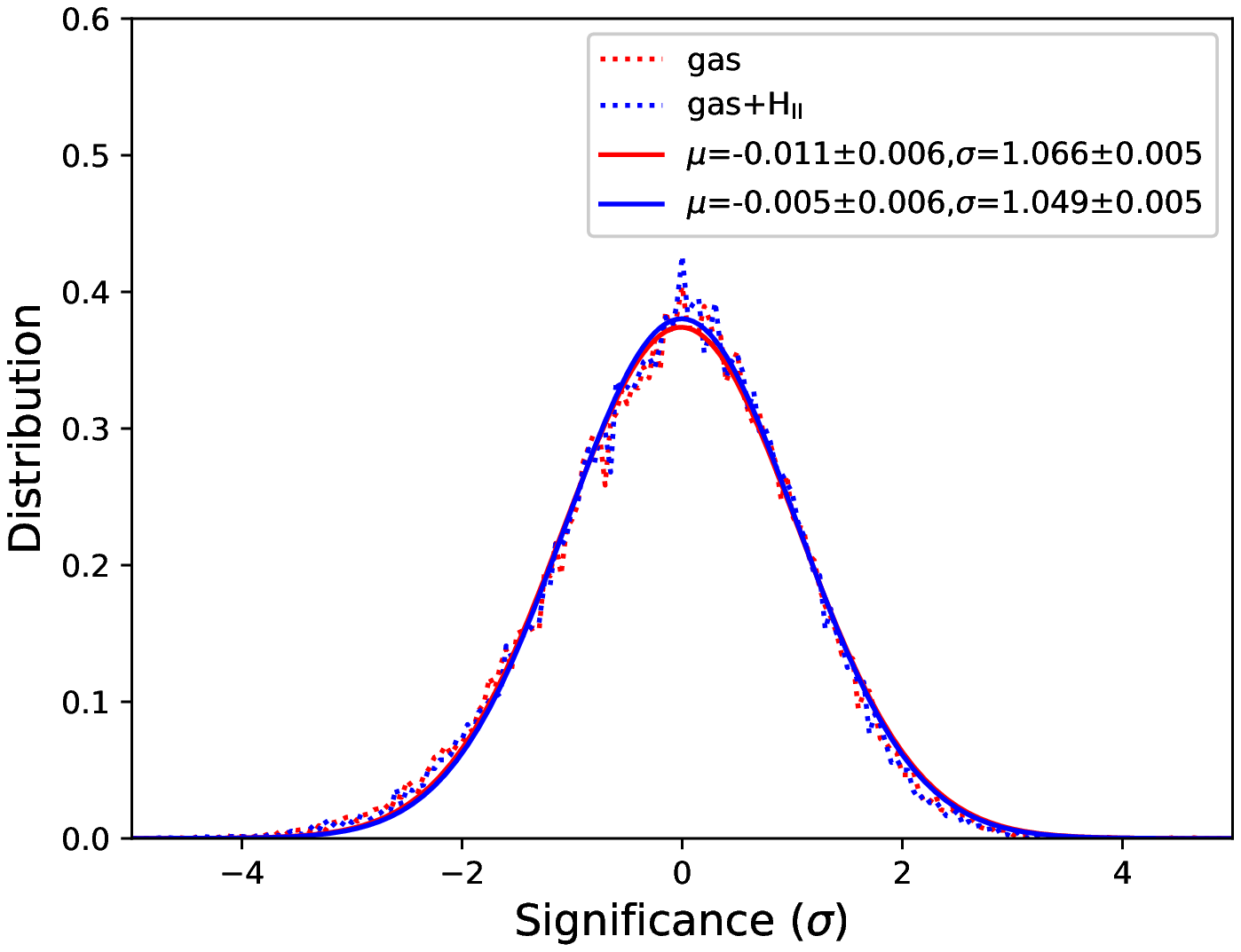}\\
\caption { Same as Fig.~\ref{fig:ts_reg1}, but for region II.} 
\label{fig:ts_reg2}
\end{figure*}

\end{appendix}

\end{document}